\documentclass[11pt]{iopart}
\usepackage[utf8]{inputenc}
\usepackage[backend=biber,style=numeric,sorting=none]{biblatex}
\DeclareUnicodeCharacter{0301}{\hspace{-1ex}\'{ }}
\usepackage[font=footnotesize,labelfont=bf]{caption}
\usepackage{sidecap}
\usepackage{dashbox}
\usepackage{hyperref}
\usepackage{lipsum}
\usepackage{subfig}
\usepackage{floatrow}
\usepackage{blindtext}
\expandafter\let\csname equation*\endcsname\relax
\expandafter\let\csname endequation*\endcsname\relax
\usepackage{amsmath}
\usepackage{float}
\usepackage{multicol}
\usepackage{braket}
\usepackage{graphicx}
\usepackage{qcircuit}
\usepackage{algorithm} 
\usepackage{algpseudocode} 
\usepackage{color}
\xyoption{color}

\newcommand\dboxed[1]{\dbox{\ensuremath{#1}}}
\usepackage{tikz}
\usepackage{mathtools}%
\usetikzlibrary{decorations.text}
\usepackage{pst-node, pst-arrow}

\usepackage{environ}
\usepackage{adjustbox}
\RequirePackage[T1]{fontenc}
\RequirePackage[utf8]{inputenc}
\RequirePackage{lmodern}
\makeatletter
\newcommand\mathcircled[1]{%
  \mathpalette\@mathcircled{#1}%
}
\newcommand\@mathcircled[2]{%
  \tikz[baseline=(math.base)] \node[draw,circle,inner sep=1pt] (math) {$\m@th#1#2$};%
}
\usetikzlibrary{
    arrows,
    calc,
    chains,
    decorations,
    decorations.text,
    decorations.pathmorphing,
    matrix,
    positioning,
    shapes,
    tikzmark,
    fit
}
\usetikzlibrary{shapes.symbols}
\usepackage{circledsteps}
\usepackage{floatrow}
\tikzset{
    mycirc/.style={circle, draw=black, thick, outer sep=0, anchor=#1, inner sep=0, minimum size=1.5ex},
    mytext/.style={text=black, anchor=#1, inner sep=0, outer sep=1pt}
}
\tikzstyle{arrow} = [thick,->,>=stealth]
\tikzstyle{decision} = [diamond, 
minimum width=2cm, 
minimum height=1cm, 
text centered, 
draw=black, 
fill=green!30]
\tikzstyle{arrow} = [thick,->,>=stealth]

\tikzset{
    mycirc/.style={circle, draw=black, thick, outer sep=0, anchor=#1, inner sep=0, minimum size=1.5ex},
    mytext/.style={text=black, anchor=#1, inner sep=0, outer sep=1pt} 
}

\usepackage{geometry}

\begin{document}

\title{Single-Step Parity Check Gate Set for Quantum Error Correction}

\author[GU, AM and SD]{Gözde Üstün$^1$, Andrea Morello$^1$ and Simon Devitt$^2$}

\address{$^1$ Centre for Quantum Computation and Communication Technology,
School of Electrical Engineering and Telecommunications,
University of New South Wales, Sydney, New South Wales 2052, Australia}
\address{$^2$ Centre for Quantum Software and Information
University of Technology Sydney, Sydney, Australia}

\begin{abstract}
A key requirement for an effective Quantum Error Correction (QEC) scheme 
is that the physical qubits have error rates below a certain threshold. 
The value of this threshold depends on the details of the specific QEC scheme, 
and its hardware-level implementation. This is especially important with parity-check circuits, which are the fundamental building blocks of QEC codes. The standard way of constructing the parity check circuit is
using a universal set of gates, namely sequential CNOT gates, single-qubit rotations and measurements. We exploit the insight
that a QEC code does not require universal logic gates, but can be simplified to
perform the sole task of error detection and correction. By building gates that
are fundamental to QEC, we can boost the 
threshold and ease the experimental demands on the physical hardware. We present a rigorous formalism for constructing and verifying the error behavior of these
gates, linking the physical measurement of a process matrix to the abstract error
models commonly used in QEC analysis. This allows experimentalists to
directly map the gates used in their systems to thresholds derived for a broad-class of QEC codes.
We give an example of these new constructions using the model system of two nuclear spins, coupled to an electron spin, showing the potential benefits of redesigning fundamental gate sets using QEC primitives, rather than traditional gate sets reliant on simple single and two-qubit gates. 
\end{abstract}
\vspace*{-5mm}
\maketitle
\section{Introduction}
\begin{figure}[H]
    \centering
   \includegraphics[width=.4\linewidth]{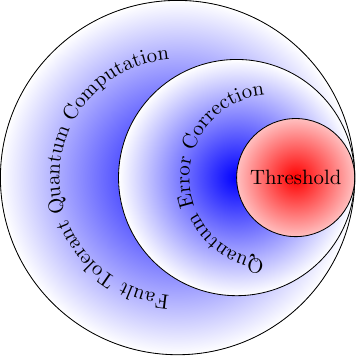}
   \caption{
   Hierarchy for scalability: The threshold is a fundamental component in practical quantum error correction (QEC), serving as a critical benchmark to gauge QEC scheme efficacy in rectifying quantum system errors. Attaining a sufficiently high threshold is essential for feasible QEC implementation; without it, practicality is compromised or rendered impossible. Crossing the threshold marks the point at which QEC becomes effective, enabling Fault Tolerant Quantum Computation (FTQC). The FTQC implementation is vital for realizing accurate and reliable quantum applications. Thus, a high threshold ensures not only QEC feasibility but also unlocks the potential for FTQC.
}
    \label{fig:hirarchy}
\end{figure}
Being able to scale quantum computers is a necessary condition for useful quantum algorithms~\cite{Scalibility}. Algorithms with proven speed up, such as Shor's factorization algorithm \cite{Shor_1997}, quantum simulation~\cite{Lanyon_2010}  or the quantum algorithm for solving linear systems of equations~\cite{Harrow_2009} only become useful when implemented on thousands of near-perfect qubits. As it is unlikely that physical error rates will be vanishingly small in the near term, the addition of Quantum Error Correction (QEC) and fault-tolerance is required. 
For practical QEC, the critical metric is the fault-tolerance threshold. This is the physical error rate below which QEC becomes effective and corrects more errors than it introduces (Figure~\ref{fig:hirarchy}). 
Therefore the threshold value, determined by the QEC scheme's structure and subsequent fault-tolerant circuit and hardware implementations~\cite{Gottesman}, is the highest acceptable physical error per gate for each qubit.


The surface code, a well-performing and experimentally feasible code~\cite{Fowler_2012, https://doi.org/10.48550/arxiv.quant-ph/9811052, Dennis_2002, Erhard_2021}, exhibits a high threshold, dependent on the assumed noise model.
A noise model incorporates assumptions about error channels like qubit loss and leakage, affecting the fault-tolerance threshold. 
Making assumptions about how error syndromes are measured in the system, commonly referred to as code capacity noise model, phenomenological noise model, and circuit-based noise model~\cite{Ashley}, can lead to significantly different fault-tolerant threshold values.
For the surface code, which involves extracting syndromes based on the parity of four-body operators, different noise models consider different methods for measuring the parity of these operators.

The code capacity and phenomenological noise models rely on natural four-body measurements, while the circuit-based model constructs four-body parity check measurements using one-body Pauli measurements and standard quantum gates. This approach yields the lowest code threshold due to physical error accumulation on standard quantum gates throughout the circuit. The circuit-based noise model is the most experimentally precise analysis of the surface code, as most experimental qubit systems lack direct access to four-body measurements.

In this study, rather than enhancing thresholds with current sources specifically designed for universal computation, we redesign the universal gates to be inherently suited for QEC implementation. Focusing on gates fundamental to QEC, as opposed to universal computation, enhances the threshold and simplifies experimental scalability challenges. This modification is deemed necessary as any large-scale quantum computer inherently functions as a QEC machine.

For error-correcting quantum computing, occasional use of traditionally universal gates suffices as primitives. However, QEC necessitates defining a new gate set, achieved by modifying the existing universal gate set. Then the challenge lies in how an experimental system can validate implementability of the new gate set in the lab, ensuring its effective use to raise the QEC threshold.

In this work, we present a rigorous formalism for constructing and verifying a native gate set, which we denote as the 
single-step parity check (SSPC) gate set, designed specifically for QEC circuits. We apply this technique to the two-body parity check circuits necessary for implementing the Honeycomb code~\cite{Gidney_2021}. We present the single-step parity check gates in the context of nuclear spin qubits in silicon~\cite{Morello_2020}, since they offer a native, direct two-body measurements via a bound electron spin \cite{M_dzik_2022}. 
However, this formalism applies more broadly and could be used for hardware systems that can naturally implement high dimensional parity check gate primitives.

The organization of this paper is as follows: Section~\ref{sec:2} briefly reviews the concept of multi-qubit measurements and introduce notations that are used for the later part of the paper. In Section~\ref{sec:4}, the Honeycomb code is explained. In Section~\ref{sec:serwan_experiment} the drawback of the circuit-based noise model is shown numerically by analyzing the experimental data~\cite{M_dzik_2022}. In Section~\ref{sec:6}, 
we introduce the Single-step parity check gate set and in Section~\ref{sec:noise}, we theoretically compare two-body parity check circuits constructed both using the universal gate set and the single-step parity check gate set in the presence of noise. 
In Section~\ref{final}, we provide an example system of silicon spin qubits to implement the SSPC gate set. In section~\ref{sec:results}, we summarise the paper.
\section{Preliminaries}
\subsection{The Concept of Making Multi Qubit Parity Check Measurement}\label{sec:2}
    Even in a single qubit, one can define the 1 state as odd parity, and the 0 state as even parity, i.e. a single qubit measurement already constitutes a parity check. In this case, it will be labelled $M_{pp1}$, where $pp$ refers to Pauli Product and $M_{pp1}$ refers to measuring in one Pauli basis, for example measuring in the $Z$ basis. 
    This can be generalized beyond a single qubit and becomes particularly useful in QEC. 
    The idea of parity checks in QEC is to have just parity information of multiple data qubits over one (ancilla) qubit without revealing any other information. In other words, we are constructing $M_{ppi}, i = 2,3,4,\cdots $ from $M_{pp1}$. To construct $M_{ppi}$ from $M_{pp1}$ for a valid quantum observable \textit{U}, we apply the controlled-U gates between the ancilla qubit and the data qubits sequentially. For example, for a parity check measurement in the \textit{X} basis, the controlled-U gates are simply CNOT gates.
    
    While any unitary can be used for the parity check, we provide an example using well-known parity check measurements in QEC. For instance, with an \textit{XX} parity check measurement ($XX = X \otimes X$), we use two CNOT gates; for an \textit{XXXX} parity check measurement ($XXXX = X \otimes X \otimes X \otimes X$), we use four CNOT gates. To measure the system over one (ancilla) qubit, we use $M_{pp1}$. \footnote{Any valid quantum observable can be used for the parity check, determined by the operator on the target side of the controlled gate.}  
    Let us look at Figure~\ref{Fig:parit_Mppa} closer to understand what a parity check does to a state. 

\begin{figure}[H]
\centering
\sidesubfloat[]{
\hspace{0.5cm}
\label{Fig:parit_Mppa}
\Qcircuit @C=1em @R=.7em {
\lstick{\ket{0}} & \gate{H} & \ctrl{1} & \ctrl{2} &  \gate{H} & \meter\\
\lstick{\ket{q_{1}}} & \qw & \targ & \qw & \qw & \qw  \\
\lstick{\ket{q_{2}}} & \qw & \qw & \targ & \qw & \qw \\
}
}
\hspace{.5cm}
\sidesubfloat[]{
\hspace{0.5cm}
\label{Mpp_b}
\Qcircuit @C=1em @R=.7em {
\lstick{ \ket{0}} & \gate{H} & \ctrl{1} & \ctrl{2} &  \ctrl{3} & \ctrl{4} &  \gate{H} & \meter\\
\lstick{ \ket{q_{1}}} & \qw & \targ & \qw & \qw & \qw & \qw & \qw  \\
\lstick{ \ket{q_{2}}} & \qw & \qw & \targ & \qw & \qw & \qw & \qw \\
\lstick{ \ket{q_{3}}} & \qw & \qw & \qw & \targ & \qw & \qw & \qw  \\
\lstick{ \ket{q_{4}}} & \qw & \qw & \qw & \qw & \targ & \qw & \qw  \\
}
}

\caption{\textit{XX} and \textit{XXXX} parity check measurements using $M_{pp1}$. a) a two-body parity check circuit. The circuit outputs the parity information of the two qubits in the $X$ basis by sequentially applying CNOT gates between the ancilla and q1 and q2, followed by measuring the ancilla -$M_{pp1}$-. b) a four-body parity check circuit. The circuit outputs the parity information of the four qubits in the $X$ basis by sequentially applying CNOT gates between the ancilla and four qubits, followed by measuring the ancilla -$M_{pp1}$-.}
\end{figure}

In Figure~\ref{Fig:parit_Mppa}, we started with an initial state of the ancilla qubit which is $\ket{0}$. The initial combined state of the system is $\ket{0}\otimes \ket{\psi_{in}}$ in which $\ket{\psi_{in}}$ is a general state over $\ket{q_1}$ and $\ket{q_2}$. Then, applying the Hadamard gate transforms the initial combined state of the system, $\ket{0}\otimes \ket{\psi_{in}}$, as follows:
\begin{equation} \label{eq:hadamard}
    \ket{0}\ket{\psi_{in}} \xrightarrow{\text{Hadamard}} \frac{1}{\sqrt{2}} \left(\ket{0}+\ket{1} \right)\ket{\psi_{in}}
\end{equation}\\
After the Hadamard gate, once we apply the CNOT gates sequentially, we have the transformation of~\ref{eq:hadamard} to~\ref{eq:CNOT}:
\begin{equation}\label{eq:CNOT}
\begin{aligned}
   \frac{1}{\sqrt{2}}(\ket{0}+\ket{1})\ket{\psi_{in}} \xrightarrow{\text{CNOT's}}  \frac{1}{\sqrt{2}}(\ket{0}\ket{\psi_{in}} + \ket{1}XX\ket{\psi_{in}})
\end{aligned}
\end{equation}
We apply the last Hadamard gate and have the transformation of~\ref{eq:CNOT} as follows:
\begin{equation}
\begin{aligned}
    \frac{1}{\sqrt{2}}(\ket{0}\ket{\psi_{in}} + \ket{1}XX\ket{\psi_{in}}) \xrightarrow{\text{Hadamard}} \\
     \frac{1}{2} \ket{0}(\ket{\psi_{in}}+XX\ket{\psi_{in}}) 
     +\frac{1}{2}\ket{1}(\ket{\psi_{in}}-XX\ket{\psi_{in}})
\end{aligned}
\end{equation}
The final step of the circuit is to measure the ancilla qubit. After this step, if:
\begin{equation}
    \begin{aligned}
      \text{ancilla} = \ket{0} \Rightarrow \frac{ \ket{\psi_{in}} + XX \ket{\psi_{in}} } {2} \cong \ket{\psi_{out}} \\
      \text{ancilla} = \ket{1} \Rightarrow \frac{ \ket{\psi_{in}} - XX \ket{\psi_{in}} } {2} \cong \ket{\psi_{out}}
    \end{aligned}
\end{equation} 
where $\cong$ means up to a renormalization factor.
The symmetries of $\ket{\psi_{out}}$ are\\
If:
\begin{equation}
\begin{aligned}
   \text{measured} \ket{0}  \Rightarrow XX \ket{\psi_{out}} &= XX \left( \frac{ \ket{\psi_{in}}+XX\ket{\psi_{in}} }{2} \right) \\
            &= \frac{XX \ket{\psi_{in}} + \ket{\psi_{in}}}{2} \\
            &= \ket{\psi_{out}}
\end{aligned}
\end{equation}
If:
\begin{equation}
\begin{aligned}
    \text{measured} \ket{1}  \Rightarrow XX\ket{\psi_{out}} &= XX \left ( \frac{\ket{\psi_{in}}-XX\ket{\psi_{in}} } {2}\right) \\
            &=  \frac{XX\ket{\psi_{in}} - \ket{\psi_{in}}}{2} \\
            &= - \left( \frac{\ket{\psi_{in}} - XX\ket{\psi_{in}}}{2} \right) \\
            &= -\ket{\psi_{out}}
\end{aligned}
\end{equation}
Hence, dependent on the measurement result, we have projected our state into eigenstates of the \textit{XX} operator with eigenvalues $+1$ or $-1$, with the ancilla measurement result dictating which eigenstate we have projected into. If the input was already in an  
eigenstate with eigenvalues $+1$ or $-1$, we will simply obtain a measurement result on the ancilla that indicates the parity of the input state, without changing the state itself.  We therefore have the parity information of the multi qubit system by using CNOT gates and a one-body measurement, ($ \text{CNOT's} + M_{pp1}$). 

Multi-body measurements are used in different aspects of quantum information/computation. In QEC, stabilizer measurements are the important examples of multi-body measurements, that occur over the operators that define the stabilizers of the QEC code. However, for example, instead of making direct \textit{XX} parity check measurement, called $M_{pp2}$, we perform this measurement by using $ \text{CNOT's} + M_{pp1}$ and single qubit rotations. Here $M_{pp2}$ refers to the fact that the operator we are measuring, \textit{XX}, is a Pauli operator with weight two.
In semiconductor spin qubits, single-triplet readout is also a multi-body measurement over 2 qubits. However, we are not using it to measure the parity of a multi-qubit Pauli operator, so we do not classify it as  $M_{pp2}$.  

Measuring a qubit is, in fact, a one-body parity measurement, allowing the unique definition of states. However, when measuring multiple qubits, our goal is to constrain the global space into subspaces where the symmetry of the measured operator is enforced. 
Consider the example in Figure \ref{Fig:parit_Mppa}, where the input state is chosen as:
\begin{equation}
   \ket{\psi_{in}} =  \alpha\ket{00}+\beta\ket{01}+\gamma\ket{10}+\delta\ket{11}
\end{equation}
where $\alpha$,$\beta$,$\gamma$ and $\delta$ are arbitrary complex amplitudes that define a unique state in this four dimensional Hilbert space. If we restrict the space by enforcing a symmetry, i.e., by running the circuits in \ref{Fig:parit_Mppa}, prior to measurement of the ancilla, the state of the three qubits before measurement, $\ket{\psi_{bm}}$, $bm$ is: 
\begin{equation} \label{Eq: XXbasisforancilla}
   \ket{\psi_{bm}} = \ket{0} \left( \frac{1}{2}(I + XX)\ket{\psi_{in}} \right) + \ket{1}\left( \frac{1}{2}(I - XX)\ket{\psi_{in}} \right).
\end{equation}
Here $\ket{0}$ and $\ket{1}$ are the states of the ancilla qubit. We know that by measuring the ancilla state will project the state into an eigenstate of $XX$.  If we assume that we measure the ancilla in $\ket{0}$ and project our state into the $+1$ eigenstate of \textit{XX}, we can write our output state $\ket{\psi_{out}}$ as
\begin{equation} \label{eq:xx_afterstate}
\begin{aligned}
  \ket{\psi_{out}} = \frac{1}{2} \left( \alpha\ket{00} + \beta\ket{01} + \gamma\ket{10}+ \delta\ket{11} \right) +
  \frac{1}{2} \left( \alpha\ket{11} + \beta\ket{10} + \gamma\ket{01} + \delta\ket{00} \right).
\end{aligned}
\end{equation}
By grouping them
\begin{equation} \label{eq:renamed}
\begin{aligned}
  \ket{\psi_{out}} = (\alpha + \delta) \left ( \frac{\ket{00} + \ket{11}} {2} \right ) + (\beta + \gamma) \left( \frac{\ket{01} + \ket{10}} {2} \right) \\
\end{aligned}
\end{equation}
we can rewrite Equation \ref{eq:renamed}  
as follows:
\begin{equation}
    \begin{aligned}
      \frac{(\alpha + \delta)}{\sqrt{2}} = \alpha^{'}  \text{and }  \frac{(\ket{00} + \ket{11})}{\sqrt{2}} = \hat{\ket{0}} \\
      \frac{(\beta + \gamma) }{\sqrt{2}}= \beta^{'}    \text{and }  \frac{(\ket{01} + \ket{10})}{\sqrt{2}} = \hat{\ket{1}}
    \end{aligned}
\end{equation}
As a result, we have:
\begin{equation} \label{xx_Cal}
   \ket{\psi_{out}} =  \alpha^{'} \hat{\ket{0}} + \beta^{'}\hat{\ket{1}}
\end{equation} which is effectively a new qubit.
Hence, by enforcing a symmetry on our original two-qubit system, we have reduced it to an effective one qubit system, where our new basis states, $\{\hat{\ket{0}},\hat{\ket{1}}\}$, are both $+1$ eigenstates of \textit{XX}. Making 
multi-body parity measurements enforces the symmetry of the system and reduces the number of degrees of freedom.  
A two-body parity measurement over a two qubit system reduces the result into effective one-qubit. This can be generalised to an $N$ qubit system: for every symmetry we enforce, by performing a measurement of an arbitrary operator, we reduce the effective number of qubits in the system by one~\cite{gottesman2009introduction}\cite{Devitt_2013}.
\subsection{Honeycomb Code}\label{sec:4}
\begin{figure}[H]
\centering
\includegraphics[width=\textwidth]{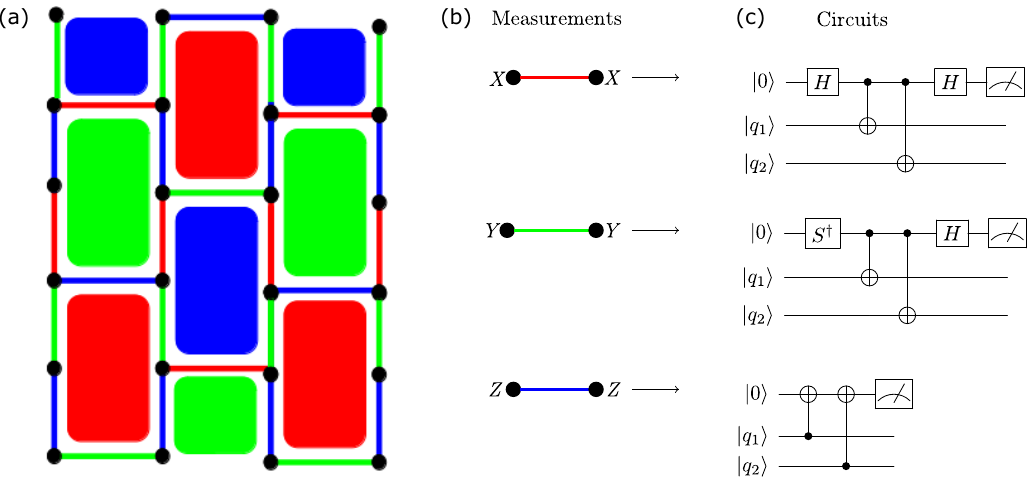}
%
\caption{Honeycomb code a) General structure of the Honeycomb code. Black dots represents physical qubits. Multiplying blue and red edges yields the green face (6 body \textit{Y} stabilizer). Multiplying blue and green edges yields the red face (6 body \textit{X} stabilizer). Multiplying red and green edges yields the blue face (6 body \textit{Z} stabilizer).  b) Measurements for the Honeycomb Code. Honeycomb code requires two body parity check: \textit{XX}, \textit{YY} and \textit{ZZ} c) Equivalent circuits for \textit{XX}, \textit{YY}, and \textit{ZZ} parity measurement using the traditional method: 1- and 2-qubit gates, plus $M_{pp1}$.
}
\label{honeycomb}
\end{figure}

In 2020, the Honeycomb code~\cite{Hastings_2021}, a variant of Floquet codes ~\cite{Floqet}, was introduced as a novel quantum error correction (QEC) code~\footnote{This code was originally introduced by Kitaev in 2006\cite{Kitaev_2006}, but its use as a QEC code was demonstrated in \cite{Hastings_2021}.} The Honeycomb code stands as one of the more recent advancements in topological code constructions.  It involves qubits positioned on a 2D lattice, forming a fixed hexagonal configuration. The distinguishing feature of the Honeycomb code, in contrast to the surface code, is its reliance on two-body parity checks, specifically $XX$, $YY$, and $ZZ$ measurements, instead of requiring $XXXX$ and $ZZZZ$ measurements. Figure~\ref{honeycomb}c provides the equivalent circuit for performing $XX$ and $ZZ$ parity check measurements in the Honeycomb code.

It was also shown that this code exhibited comparable thresholds to the surface code~\cite{Wang_2011}, in contrast to Gauge codes which can be realised using two-body parity checks, but whose stabilisers are generally much higher weights~\cite{Bombin_2010}. Since the Honeycomb code requires only two-body parity measurements, $M_{pp2}$, which may be achievable experimentally in some systems naturally.

In~\cite{Gidney_2021}, Gidney and Fowler demonstrated that when considering a scenario where a natural $M_{pp2}$ measurement is available from the hardware, the error modeling mirrors a phenomenological noise model. This involves applying a perfect multi-qubit measurement followed by error mapping according to the equation:
\begin{equation} \label{eq:Mpp2}
    \{I,X,Y,Z\}^{\otimes 2} \times \{flip, no flip\}
\end{equation}
with probability, $p$. This error map allows us to differentiate between two types of errors: quantum errors and classical errors. The first set of brackets represents the quantum errors and it allows us to have one Pauli error on one of the data qubits. The second set of brackets represents the classical error and it allows us to have one classical bit flip error which is the bit flip error on the classical measurement result on the ancilla qubit. With this map, the threshold of the Honeycomb code was found to be approximately 2\%. This threshold is commensurate with the phenomenological threshold for the surface code, but unlike the surface code, some hardware systems may have the possibility of realising $M_{pp2}$ operations, while it is unlikely that a natural $M_{pp4}$ measurement will be found in any current hardware systems. 
 
This means that if we can have a natural two-body measurement that satisfies the error model of Gidney and Fowler, we will have a three times higher threshold for the Honeycomb code while maintaining a 2D Nearest-Neighbour hardware architecture. Furthermore, the latest study by Gidney~\cite{Gidney_2022}, shows that if the surface code is run by using natural two-body measurements, the threshold of the surface code will be similarly increased. 

The next question is: how can we asses whether a physical system allows a natural implementation of natural $M_{pp2}$ operations? For this purpose, we outline a workflow to map an experimental process matrix to a QEC-specific error model for checking if a real system's error behavior matches the assumptions of a threshold analysis. 
We develop and validate a native gate set tailored for QEC, focusing on higher dimensional unitaries required for syndrome extraction circuits, termed the single-step parity check gate set. Applying this technique to two-qubit parity check circuits for the Honeycomb code implementation, we illustrate the single-step parity check gates in the context of spin qubit systems \cite{morello2020donor,M_dzik_2022}.

\section{Analyzing the XX Parity Check Circuit from Experimental Data in the Context of QEC}\label{sec:serwan_experiment}
Assessments of hardware systems have primarily focused on single- and two-qubit gate fidelities, assuming fidelity reflects operational perfection. However, fidelity is a numerical representation lacking information on error type, and it doesn't correspond to gate error rates. 
In QEC, we analyze errors using stochastic Pauli channels, focusing on error types and values for 1- and 2-qubit gates instead of relying on gate fidelities. Furthermore, relying solely on the error values for these gates is insufficient to determine the total circuit error rate and, consequently, the overall circuit perfection rate due to error propagation in the circuit.

Here we numerically demonstrate (to the best of our knowledge, for the first time) the loss in the circuit by calculating the perfection rate at each step when utilizing one-body measurement ($M_{pp1}$) with 1- and 2-qubit gates, based on experimental data. 
The detailed explanation of the perfection rate calculation, including a comprehensive tutorial~\ref{Mpp1}, is available in the appendices. Our algorithm~\ref{algorithm} 
allows experimentalists to directly map the gates used in their systems to thresholds derived for a broad-class of QEC codes.

\begin{figure}[!htb]
      \centering
      \includegraphics[width=.7\textwidth]{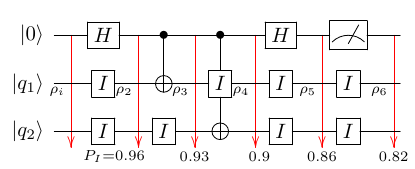}
        \caption{Analysis of the \textit{XX} parity check circuit in terms of QEC.  The circuit depth is 6 (6 steps to calculate in the widget). Commencing with $\rho_i$, each step the value of having no error in the channel ($P_I$) decrease since the errors accumulate to the next step of the circuit 
        }
        \label{fig:xx}
    \end{figure}

In this section, we use data from~\cite{M_dzik_2022} to show how the output perfection rate decreases for $XX$ parity check circuit which is the smallest parity check measurement that we can use to implement a topological error correcting code. This study showcases the universal quantum logic operations performed on a silicon nanoelectronic device utilizing a pair of ion-implanted $^{31}$P donor nuclei. This device, consisting of two phosphorus nuclei and one electron spin system, is henceforth referred to as the `2P1e' device. Researchers precisely characterized the quantum operations by using Gate Set Tomography~\cite{Greenbaum_2015}\cite{Nielsen_2021}. They achieved one-qubit gates with average gate fidelities up to 99.95(2)\%,  two-qubit gates with an average gate fidelity of 99.37(11)\% and two-qubit preparation/measurement fidelities of 98.95(4)\%. These results indicate that nuclear spins in silicon are close to the performance required by fault-tolerant quantum processors. 
Figure~\ref{fig:xx} shows the decrease of perfection rate for the \textit{XX} parity check measurement which is constructed by the circuit based model. Another important point is the identity gates. When we apply a single qubit gate on a multi-qubit circuit, the qubits which have no applied gate should stay stable and this is represented by identity gates. Identity gates are also as important as other single-qubit gates since they contribute to the errors.
As each step progresses in Figure~\ref{fig:xx}, errors accumulate. Initially, the system is in state $\rho_i$. Then, a Hadamard gate is applied to the first qubit and an identity is applied to the other two, resulting in state $\rho_2$ from $\rho_i$. The final density matrix for $\rho_2$ is calculated,  in terms it of having experienced certain errors in a QEC block, resulting in a perfection rate of 0.96.

The next step, $\rho_3$, includes errors from not only CNOT and identity operations, but also from $\rho_2$. When the process continues, the final perfection rate is found to be 0.82. The perfection rate between the application of the first operation - in this case, the Hadamard gate on the ancilla qubit and the identities on the $\ket{q_1}$ and $\ket{q_2}$ - and the measurement operation decreased by 14\%, from 0.96 to 0.82. This striking observation highlights how, even in a state-of-the-art experiment with average gate fidelities above 99\%, constructing a two-body parity check circuit out of universal logic gates results in a drastic reduction of the perfection rate at the end of circuit. 

With the goal of optimizing the perfection rate for an $M_{pp2}$ operation, we thus aim to design full unitaries directly, instead of composing them from 1- and 2-qubit gates. 
We construct these new unitaries by modifying the universal gate set.
 These unitaries are fundamental to QEC rather than universal computation. By using these gates, we can increase the threshold and decrease the experimental challenge to the scalibility. We call these gates as single step-parity check gate set. These gates are actually multi-body parity measurement gates. This is described in section~\ref{sec:6}



\renewcommand{\algorithmicrequire}{\textbf{Input:}}
\renewcommand{\algorithmicensure}{\textbf{Output:}}
\newcommand{\algorithmicbreak}{\textbf{break}}
\newcommand{\BREAK}{\STATE \algorithmicbreak}
\captionsetup[algorithm]{labelformat=empty} 
\newcounter{phase}[algorithm]
\newlength{\phaserulewidth}
\newcommand{\phaseTitle}{Phase} 
\newcommand{\setphaserulewidth}{\setlength{\phaserulewidth}}
\newcommand{\phase}[1]{%
  \vspace{-1.25ex}
  \Statex\leavevmode\llap{\rule{\dimexpr\labelwidth+\labelsep}{\phaserulewidth}}\rule{\linewidth}{\phaserulewidth}
  \Statex\strut\refstepcounter{phase}\item[\textbf{\phaseTitle~\thephase:~}\textit{~#1}] 
  \vspace{-1.25ex}\Statex\leavevmode\llap{\rule{\dimexpr\labelwidth+\labelsep}{\phaserulewidth}}\rule{\linewidth}{\phaserulewidth}}

\section{Single-Step Parity Check Gate Set and \texorpdfstring{$M_{pp2}$}{M\_{pp2}} } \label{sec:6}
\begin{figure}[H]
\centering
\resizebox{\textwidth}{!}{%
\sidesubfloat[]{
\hspace{0.5cm}
\centering
\Qcircuit @C=1em @R=2em {
\lstick{ \ket{0}} & \gate{H} & \ctrl{1} & \ctrl{2} &  \gate{H} & \meter\\
\lstick{ \ket{q_{1}}} & \qw & \targ & \qw & \qw & \qw  \\
\lstick{ \ket{q_{2}}} & \qw & \qw & \targ & \qw & \qw \gategroup{1}{2}{3}{5}{.7em}{--}
}

\label{fig:mpp2circuit}
}
\sidesubfloat[]{
\hspace{0.5cm}
\centering
\Qcircuit @C=1em @R=2em {
\lstick{ \ket{0}} & \multigate{2}{\{I,X,Y,Z\}^{\otimes2}} & \meter\\
\lstick{ \ket{q_{1}}} & \ghost{\{I,X,Y,Z\}^{\otimes2}}& \qw \\
\lstick{ \ket{q_{2}}} & \ghost{\{I,X,Y,Z\}^{\otimes2}}& \qw
}
\label{fid:mpp2circuitb}
}
\sidesubfloat[]{
\hspace{0.5cm}
\centering
\Qcircuit @C=1em @R=2em {
\lstick{ \ket{0}} & \multigate{2}{U_{XX}} & \meter\\
\lstick{ \ket{q_{1}}} & \ghost{U_{XX}}& \qw \\
\lstick{ \ket{q_{2}}} & \ghost{U_{XX}}& \qw
}
\label{fig:mpp2circuitc}
}
}
\caption{Single-step Parity Check Gate for \textit{XX}. a) $M_{pp2}$ with current resources. The gates inside the dashed rectangle represent the 1- and 2-qubit operations normally used to build the \textit{XX} $M_{pp(2)}$.  b) A natural $M_{pp(2)}$. The noise model incorporates Pauli errors from the set defined in Equation Equation~\ref{eq:Mpp2}. c) Natural $M_{pp(2)}$ as a widget: new unitary $U_{XX}$ and 1-qubit measurement operator. }
\end{figure}
\begin{figure}[H]
\centering
\sidesubfloat[]{
\hspace{0.5cm}
\centering
\Qcircuit @C=1em @R=2em {
\lstick{ \ket{0}} & \targ & \targ  & \meter\\
\lstick{ \ket{q_{1}}} & \ctrl{-1} & \qw & \qw & \qw  \\
\lstick{ \ket{q_{2}}} & \qw & \ctrl{-2} & \qw & \qw \gategroup{1}{2}{3}{3}{.7em}{--}
}
\label{fig:mpp2zzcircuit}
}
\sidesubfloat[]{
\hspace{0.5cm}
\centering
\Qcircuit @C=1em @R=2em {
\lstick{ \ket{0}} & \multigate{2}{\{I,X,Y,Z\}^{\otimes2}} & \meter\\
\lstick{ \ket{q_{1}}} & \ghost{\{I,X,Y,Z\}^{\otimes2}}& \qw \\
\lstick{ \ket{q_{2}}} & \ghost{\{I,X,Y,Z\}^{\otimes2}}& \qw
}
\label{fig:mpp2zzcircuitb}
}
\sidesubfloat[]{
\hspace{0.5cm}
\centering
\Qcircuit @C=1em @R=2em {
\lstick{ \ket{0}} & \multigate{2}{U_{ZZ}} & \meter\\
\lstick{ \ket{q_{1}}} & \ghost{U_{XX}}& \qw \\
\lstick{ \ket{q_{2}}} & \ghost{U_{XX}}& \qw
}
\label{fig:mpp2zzcircuitc}
}
\caption{Single-step Parity Check Gate for \textit{ZZ}. a) $M_{pp2}$ with current resources. The gates inside the dashed rectangle represent the 2-qubit operations normally used to build the \textit{ZZ} $M_{pp(2)}$.  b) A natural $M_{pp(2)}$. The noise model incorporates Pauli errors from the set defined in Equation Equation~\ref{eq:Mpp2}. The error model for both XX and ZZ parity checks is the same.
c) Natural $M_{pp(2)}$ as a widget: new unitary $U_{ZZ}$ and 1-qubit measurement operator.}
\end{figure}
Single-step parity check gates are a version of pulse optimization where we optimize for the full unitary needed for QEC. In this case, we reduce the number of steps and instead of having six different steps, we have single step for the same circuit. From Figure~\ref{fig:mpp2circuit} and \ref{fig:mpp2zzcircuit}, the unitaries are found to be Equation~\ref{unitary_xx} and Equation~\ref{unitary_zz} and they belong to the Clifford group. Apart from the new unitary, we have a new device which can be called a natural $M_{pp2}$. It is a widget that includes one of those new unitaries, and a measurement operator. This natural $M_{pp2}$ corresponds to Equation~\ref{eq:Mpp2}. It receives 3 qubits as input, using one of them as an ancilla, and returns the parity information by measuring the ancilla qubit. In the next section, we compare the single-step parity check gate and the decomposed parity check circuit theoretically in the presence of noise.
\begin{equation} \label{unitary_xx}
    \begin{aligned}
    U_{XX}=\frac{1}{2}\begin{pmatrix}
      1 & 0 &  0  &  1 & 1 &  0  &  0  & -1\\ 
      0 & 1&  1 &  0  & 0  &  1 & -1 &  0\\
      0 & 1&  1 &  0  & 0  & -1 & 1  &  0\\
      1 & 0 &  0  &  1 &-1 &  0  & 0   &  1\\
      1 & 0 &  0  & -1 & 1 &  0  & 0   &  1\\
      0 & 1& -1 &  0  & 0  &  1 & 1  &  0\\
      0 &-1&  1 &  0  & 0  &  1  &1  & 0\\
     -1&  0&  0  &  1 & 1 &  0   &0   &1
\end{pmatrix}
    \end{aligned}
\end{equation}
\begin{equation} \label{unitary_zz}
    \begin{aligned}
    U_{ZZ}=\begin{pmatrix}
      1 & 0 & 0 & 0 & 0 & 0 & 0 & 0\\
      0 & 0 & 0 & 0 & 0 & 1 & 0 & 0\\
      0 & 0 & 0 & 0 & 0 & 0 & 1 & 0\\
      0 & 0 & 0 & 1 & 0 & 0 & 0 & 0\\
      0 & 0 & 0 & 0 & 1 & 0 & 0 & 0\\
      0 & 1 & 0 & 0 & 0 & 0 & 0 & 0\\
      0 & 0 & 1 & 0 & 0 & 0 & 0 & 0\\
      0 & 0 & 0 & 0 & 0 & 0 & 0 & 1
\end{pmatrix}
    \end{aligned}
\end{equation}
\subsection{
Comparison of decomposed XX parity check circuit and XX-SSPC circuit in the presence of noise
}\label{sec:noise}
\begin{figure}[!htb]
\centering
\includegraphics[width=.65\textwidth]{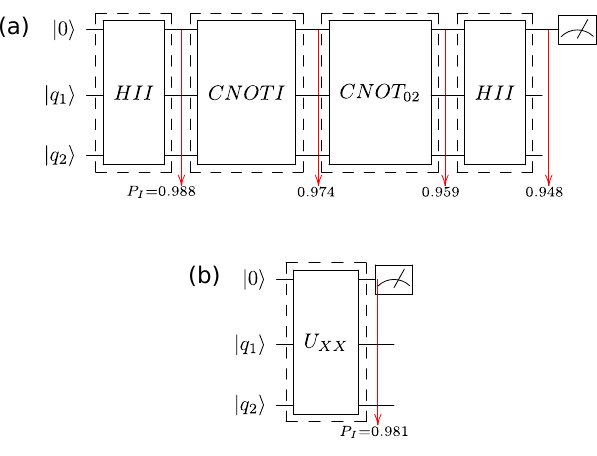}
\caption{The circuits are modeled using a standard Pauli-based noise model, where errors are randomly chosen from possible combinations of Pauli operators for all gates, representing phase flip errors applied with probability p. In this case, the probability of having, for example, Pauli Z errors includes combinations such as $Z\otimes I\otimes I$,$I\otimes Z\otimes I$, and $I\otimes I\otimes Z$ with an error rate of $p$, $Z\otimes Z\otimes I$, $I\otimes Z\otimes Z$, and $Z\otimes I\otimes Z$ with an error rate of $p^2$, and finally $Z\otimes Z\otimes Z$ with an error rate of $p^3$ for $HII$. a) A traditional, decomposed, \textit{XX} parity check circuit. This circuit comprises four building blocks: $HII$, $CNOTI$, $CNOT_{02}$, and again $HII$, constructed from $H$, $I$, and $CNOT$ gates.  
b) Defines a direct \textit{XX} parity check gate with the $XX$-SSPC gate. We create each of building block with $\approx 99\%$ fidelity.}
\label{fig:10}
\end{figure}

In this section, we will show how the perfection rate differs in the presence of noise between decomposed parity check circuit and single-step parity check circuit. For this purpose, we take the circuits defined in Figure~\ref{fig:10}. In Figure~\ref{fig:10} a, the circuit has four building blocks constructed from Hadamard, Identity, and CNOT gates: $ HII = H\otimes I \otimes I$, $CNOTI = CNOT \otimes I$, $CNOT_{02}$ (CNOT gate between the ancilla and the second data qubit), and $HII = H \otimes I \otimes I$. In Figure~\ref{fig:10}b the circuit contains only one building block, the $U_{XX}$ gate. Each noisy operator can be modelled by perfect operator plus some errors to be mapped in the circuit. 
We simulated process matrices such that each gate is modeled by a noisy channel with a 0.00375 phase flip error rate. Then we calculate the average gate fidelity from these noisy channels and we find that each building block in the circuit will have $\approx 99\%$ fidelity. Note that if we change the error type, the amount of error required to maintain the same fidelity will also change. The reason for selecting this error type is because phase flip errors are the predominant sources of error for most solid-state hardware systems, 
especially in semiconductor spin qubits such as the one described in~\cite{M_dzik_2022}.
Overall, we have on one side four building block with 99\% fidelity, and on the other we have one building block with 99\% fidelity. We apply our workflow~\ref{Mpp1}to find the perfection rate of each circuit defined in Figure~\ref{fig:10}.

As it is seen from the result, since the traditional, decomposed, parity check circuit requires four steps (four building blocks), the errors in each step accumulate. However, with a SSPC gate, there is no more than one building block so there is no place for errors to accumulate. For this reason, the perfection rate remains high.

\begin{figure}[H]
\centering
\includegraphics[width=.55\textwidth]{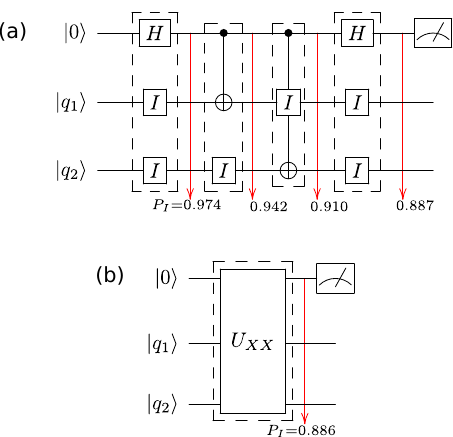}
\caption{This example demonstrates increasing the error threshold by reducing circuit gates. Unlike the example in Figure~\ref{fig:10} example, here we create two circuits with different error rates to achieve approximately the same perfection rate. a) A traditional \textit{XX} parity check circuit with Hadamard and identity gates as single-qubit gates, and CNOT gates as two-qubit gates. The circuit is  modelled using a standard Pauli based noise model where errors are randomly chosen over X, Y and Z for single qubit gates (including idling gates, initialisation and measurement) and over the 15 possible combination of two-qubit Pauli operators for all two qubit gates, applied with a probability. Each single-qubit gate is created with $p=0.0085$ (corresponding to 99.43\% fidelity). 
Errors calculated in stochastic Pauli channels yield a total perfection rate of $P_I=0.887$ at the circuit's end. b) Using a direct XX parity check gate with an \textit{XX}-SSPC gate, we can tolerate a much larger phase-flip error rate per qubit, 
$p=0.0239$ to achieve a similar perfection rate as in (a). 
Hence, the noise model takes Pauli errors from the set defined in Equation~\ref{eq:Mpp2}. 
}
\label{fig:11}
\end{figure}

As a further illustration, in Figure~\ref{fig:11} we show a simulation where each single qubit gate (Hadamard and Identity gates) has 99.43\% fidelity, and the CNOT gate experiences the single qubit errors as a tensor product of the amount ($I \otimes X$,$I \otimes Y$...). We chose this fidelity value because it is the surface code fault-tolerance threshold~\cite{Fowler_2012}. To achieve this fidelity, we used an error rate of $p = 0.0085$ for the phase flip error in each single-qubit gate. As depicted in Figure~\ref{fig:11}, the resulting perfection rate at the circuit's end is 0.887. This perfection rate is obtained by utilizing gates with the specified fault tolerance error value.
 

To attain an equivalent perfection rate for the XX-SSPC gate, comparable to the decomposed parity check circuit, we can increase the error rate to $p = 0.0239$ for the phase flip error per qubit. Implementing an XX gate with this error rate yields a perfection rate $P_I = 0.886$, as illustrated in Figure~\ref{fig:11}b. 
As a result once the parity check circuit is created with SSPC gate at higher error rates than the 1- and 2-qubit gates, the same perfection rate can be maintained. 
This simple example shows how SSPC gates can drastically boost the error threshold value for QEC.
\subsection{Practical implementation of SSPC Gates}
The theoretical discussion in the preceding section illustrates the advantages afforded by implementing parity checks using a single-step $M_{pp2}$ operation instead of breaking it down to a sequence of individual gates. We now illustrate a practical example of how a SSPC gate could be implemented in a real physical system. 

\begin{figure}[!htb]
    \centering
    \includegraphics[width=\linewidth]{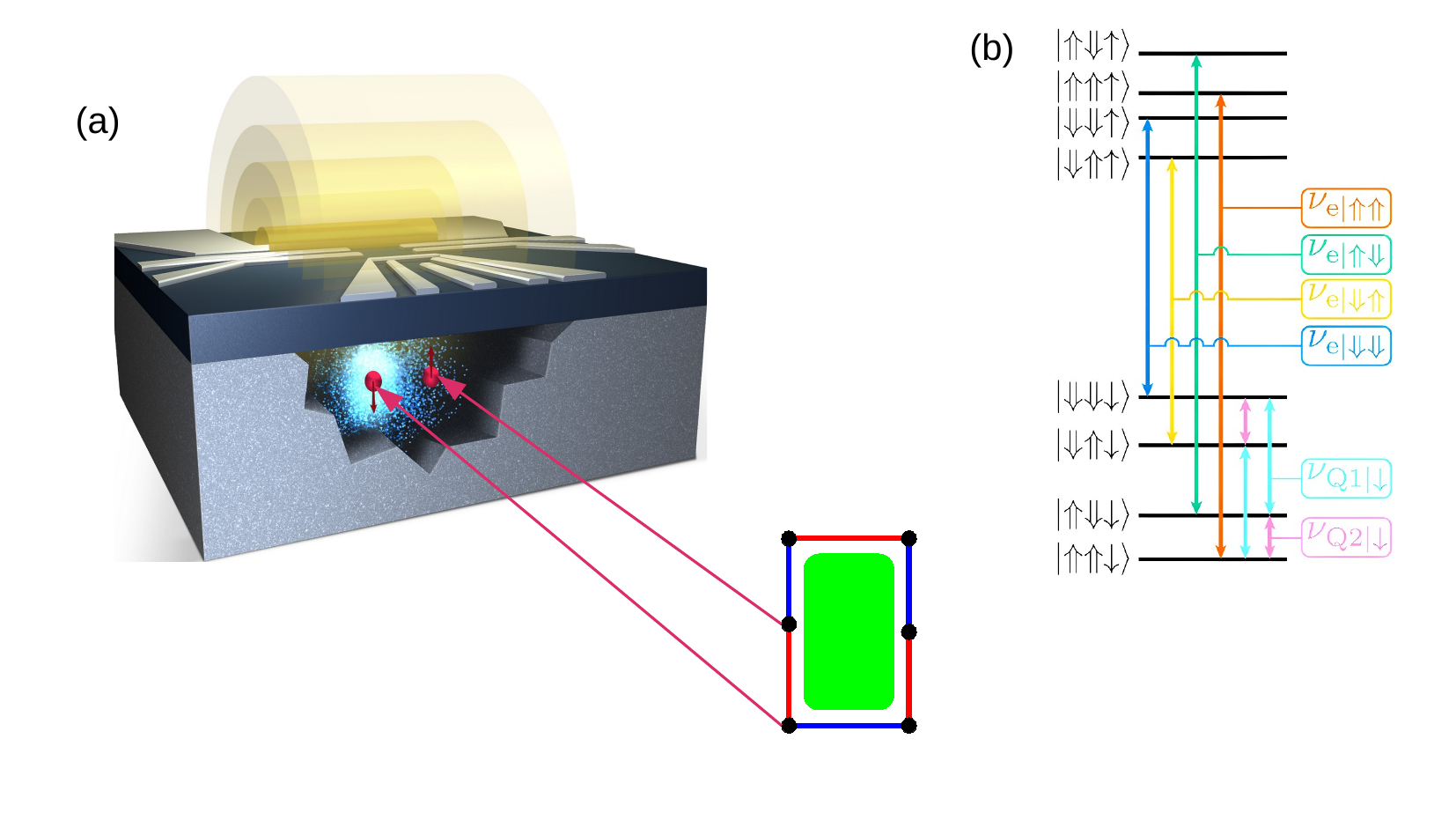} 
    \caption{a) The left side of the image is an artist's representation of the 2P1e device~\cite{M_dzik_2022}, while the right side corresponds to the smallest honeycomb code cell. The two black dots in the honeycomb code represent the two phosphorus nuclear spins that are physical qubits in the 2P1e device, shown in pink. The ancilla qubit is an electron spin (blue 'cloud') hyperfine-coupled to both nuclei. With the 2P1e system, we are able to demonstrate the smallest parity check circuit, which is the smallest component of the Honeycomb code cell. b) NMR and ESR transitions in the eight-dimensional Hilbert space of the two-phosphorus system when loaded with an electron. The pink and light blue lines represent the NMR frequencies, while the remaining lines represent the ESR frequencies. Adapted from Ref.~\cite{M_dzik_2022}}
    \label{fig:2p1e_honeycomb}
\end{figure}

Let us consider a three-spin system consisting of an electron spin, acting as the ancilla, and two nuclear spins, acting as the qubits. Their drift Hamiltonian in the laboratory frame takes the general form:
\begin{equation} \label{eq:Hamiltonian}
    \Vec{H} = - \gamma_e B_0 \hat{S_z} - \gamma_n B_0 (\hat{I_{z1}}+\hat{I_{z2}}) + A_1 \Vec{S}\cdot\Vec{I_1} + A_2 \Vec{S}\cdot\Vec{I_2},
\end{equation}
where $\Vec{S} = [\hat{S_x},\hat{S_y},\hat{S_z}]$ are the electron spin operators with eigenvectors $\{\ket{\uparrow},\ket{\downarrow}\}$, $\Vec{I_i} = [\hat{I_x},\hat{I_y},\hat{I_z}]$ are the nuclear spin operators for nucleus $i \in 1,2$ with eigenvectors $\{\ket{\Uparrow_i},\ket{\Downarrow_i}\}$, $A_1$ and $A_2$ are the electron-nuclear hyperfine interactions, $\gamma_e$ is the electron gyromagnetic ratio, $\gamma_n$ is the nuclear gyromagnetic ratio, and $B_0$ is a static magnetic field oriented along the $\hat{z}$-axis. This Hamiltonian is commonly found in well-studied spin qubit systems such as the nitrogen-vacancy (NV) center in diamond \cite{waldherr2014quantum}, spins in quantum dots \cite{hensen2020silicon} and donors in silicon \cite{morello2020donor}. A Hamiltonian with this structure can be viewed as implementing the smallest possible parity check circuit for the Honeycomb Code, as shown in Figure~\ref{fig:2p1e_honeycomb}. 

We take the specific example of a recent experiment where two $^{31}$P nuclear spins were hyperfine-coupled to a common electron (2P1e) and entangled using a geometric CZ gate obtained through a rotation of the electron conditional on the state of the nuclei \cite{M_dzik_2022}. The gate fidelities extracted from GST were used in the preceding section to illustrate how gate errors build through a circuit. The numerical parameters of the Hamiltonian are $\gamma_e = -27.97$~GHz/T, $\gamma_n = 17.23$~MHz/T, $B_0 = 1.33$~T, $A_1 = 95$~MHz and $A_2 = 9$~MHz. The large separation of energies in the system, whereby the electron Zeeman energy splitting $\gamma_e B_0 \approx 37.2$~GHz is orders of magnitude larger than both the nuclear Zeeman energies $\gamma_n B_0\approx 22.8$~MHz and the hyperfine interactions, ensures that the eigenstates of the Hamiltonian are almost exactly the tensor products $\{\ket{\uparrow},\ket{\downarrow}\}\otimes\{\ket{\Uparrow_1},\ket{\Downarrow_1}\}\otimes \{\ket{\Uparrow_2},\ket{\Downarrow_2}\}$, as shown in Figure~\ref{fig:2p1e_honeycomb}b. 

To implement the \textit{ZZ}-SSPC gate we add the control Hamiltonian:
\begin{equation} \label{control}
    H_{rf}(t) = - \gamma_e \Vec{B_1} \Vec{S}\cos(\omega t) - \gamma_n \Vec{B_1} (\Vec{I_1}+\Vec{I_2})\cos(\omega t)
\end{equation}
where $B_1$ is an the oscillating magnetic field strength oriented along $\hat{y}$. A \textit{ZZ}-SSPC gate is obtained trivially by applying a bichromatic pulse of $B_1$ at the frequencies $\omega_{\Uparrow\Downarrow}$ and $\omega_{\Downarrow\Uparrow}$ corresponding to the resonance frequencies of the electron spin when the nuclei are in the odd-parity states $\ket{\Uparrow\Downarrow}, \ket{\Downarrow\Uparrow}$. This is depicted with the yellow and green lines in Figure~\ref{fig:2p1e_honeycomb}b. Calibrating the amplitude and duration of the pulse such that the electron undergoes an exact $\pi$-rotation implements the CNOT gates depicted in Figure~\ref{fig:mpp2zzcircuitc}(a). The two frequencies can be applied simultaneously because they both act on the same physical object (the electron). The SSPC gate is then followed by electron readout \cite{Morello_2010}.

\label{final}
The $XX$-SSPC gate, shown in Fig.~\ref{fig:mpp2circuitc}(a), involves operations on both the electron and the nuclei, which require different time scales due to the vastly different gyromagnetic ratios. It is no longer immediately obvious that the whole operation can be executed in a single step. In order to verify whether an $XX$-SSPC gate exists, we adopt a widespread quantum control optimization algorithm known as GRAPE (Gradient Ascent Pulse Engineering)~\cite{GRAPE}, which can be found within QuTip~\cite{Li_2022}. One of the reasons we chose to use GRAPE is because it enables us to account for real-life experimental limitations such as the maximum and minimum control amplitudes. Another reason is that the GRAPE algorithm was originally designed for NMR pulses~\cite{GRAPE}\cite{Ryan_2008} and finds practical application in many experimental studies~\cite{Koch_2022}. Notably, the results of the first experiment in quantum error correction in 1998~\cite{Cory_1998} were greatly improved through the utilization of the GRAPE algorithm in 2011~\cite{Zhang_2011}.
 The GRAPE algorithm in QuTip, is wrapped by the L-BFGS-B~\cite{bfgs}\cite{Byrd_9195} optimization algorithm which takes into account second order derivatives and chooses the step size which is the change of position in the control landscape by itself. This helps the algorithm to use a smaller number of iterations. We utilized this algorithm in two distinct ways, namely modulated GRAPE and non-modulated GRAPE. The non-modulated GRAPE does not use the modulated control Hamiltonian, but it does use a control Hamiltonian with proper frequencies for both the electron and nucleus and control fields applied in two different directions ($x$ and $y$). The modulated GRAPE, however, allows us to use a modulated control Hamiltonian, and we refer to this version of the algorithm as modulated GRAPE.
 
 For the modulated GRAPE method, we took into account the $\cos(\omega t)$ term and we modulated the control Hamiltonian with $\cos(\omega t)$ for each time slot with a proper angular frequency. 
As a first step, we set the maximum control amplitude (the $B_1$ magnetic field in our case) in the range of millitesla and we ran the algorithm for 1, 10, 100, 1000 microseconds evolution time for the  
$XX$ gate. The number of time slots is one of the most important parameters of the algorithm. We need to provide a fine enough resolution to observe the oscillation of the pulse. Since we are in the laboratory frame, the time interval is determined by the highest frequency in the system, which in this case is the electron Larmor frequency. In a 1.33 T magnetic field, the electron precesses at 27.97 GHz, corresponding to a period of 35 ps. When the number of sampling per oscillation is taken into account, for each evolution time, the minimum time resolution is maintained at 10 picoseconds by allocating the appropriate number of time slots. To execute the algorithm with such a small time resolution, we utilized a High-Performance Computing (HPC) cluster. 
As a result, we could have  perfect gate fidelity for an evolution time of order 10~$\mu$s. 

\begin{figure}[!htb]
    \centering
    \includegraphics[width=.7\textwidth]{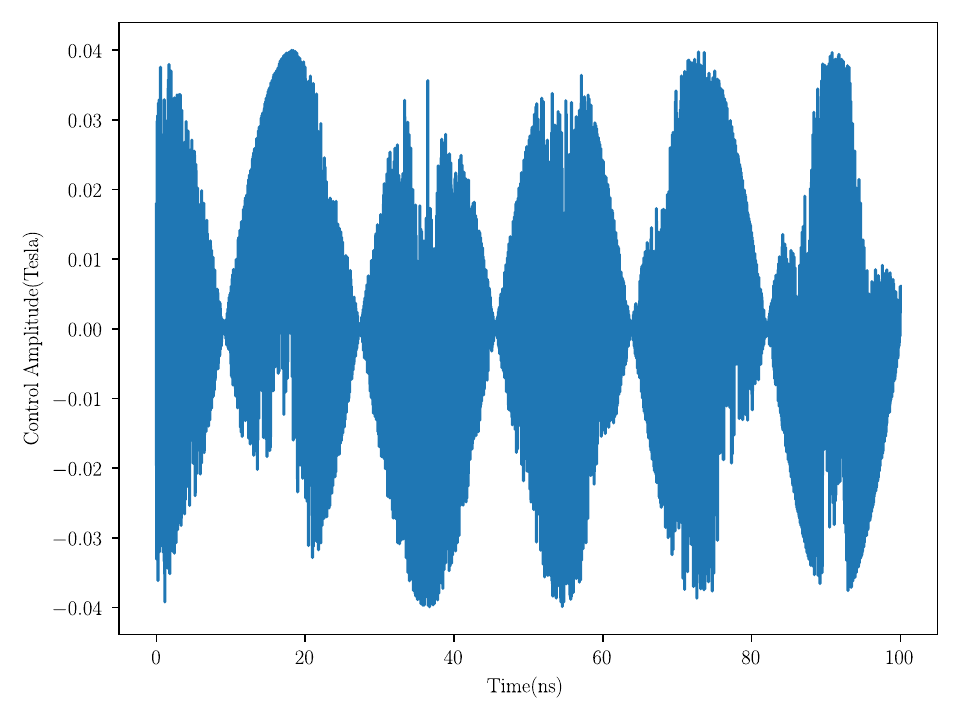} 
    \caption{The first 100 ns of the optimised control pulses for the \textit{XX} gate. The control amplitude represents the $B_1$ magnetic field applied to the spins. }
    \label{fig:pulse}
\end{figure}

We then attempted to find a shorter gate evolution time, by increasing the control amplitude (still within experimentally feasible values).
We set the evolution time to 4 microseconds for the \textit{XX} gate. We set upper and lower bounds to 40 mT for the control field and we used 400,000 time slots which correspond to every 10 picoseconds for a pulse. We run the algorithm for every frequency in the system and the algorithm was able to find \textit{XX} parity check gates with a single pulse with an accuracy of 0.9999\footnote{This is the theoretical fidelity and this demonstrates the feasibility of implementing this gate under the given conditions.} 
The first 100 ns of the control amplitudes are shown in Figure~\ref{fig:pulse}.

\begin{figure}[!htb]
    \centering
    \includegraphics[width=.7\textwidth]{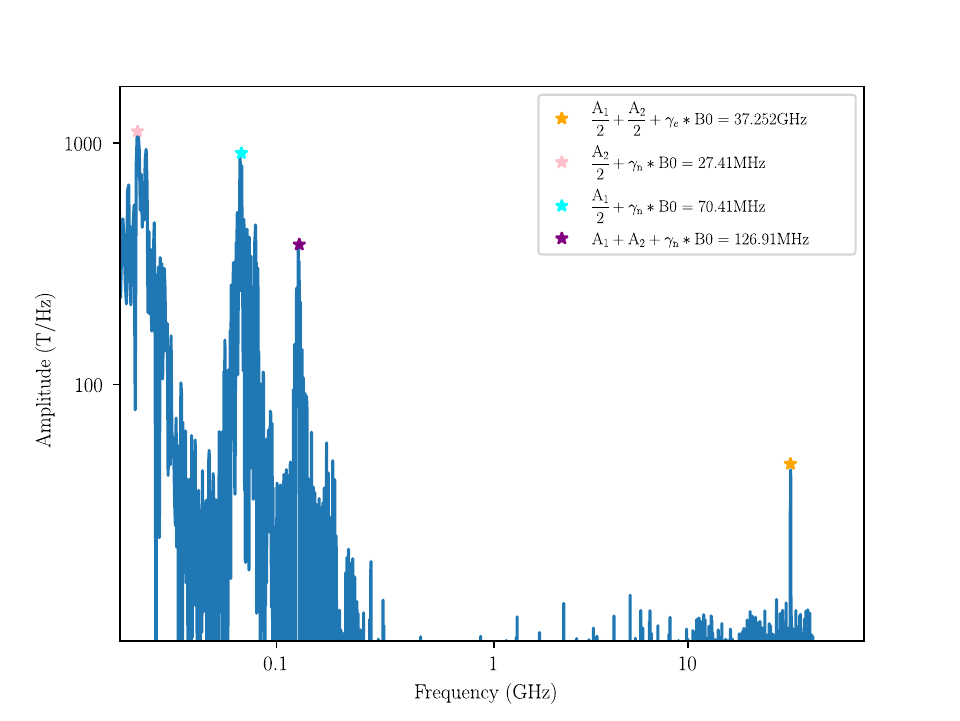} 
    \caption{Fourier Transform of the GRAPE Pulse that implements the $XX$-SSPC gate. The peaks in the spectrum correspond to the natural resonance frequencies of the electron and the nuclei. The colors of the stars represent the transitions, which are drawn with the same color as in Figure~\ref{fig:2p1e_honeycomb} }
    \label{fig:fft}
\end{figure}

To verify that the pulse found by GRAPE was physically meaningful, we then calculated its fast Fourier transform (FFT), shown in Figure~\ref{fig:pulse}.
Since the \textit{XX} gate is an entangling gate formed through a combination of operations on both nuclei and electrons, we expect the spectrum of the pulse to consists of control signals at the natural resonance frequencies of the electron (ESR) and the nuclei (NMR).
Indeed, the FFT in Figure~\ref{fig:fft} reveals all the expected frequencies: electron spin resonance (orange marker in Fig.~\ref{fig:fft}, corresponding to the orange line in Fig.~\ref{fig:2p1e_honeycomb}b), and nuclear magnetic resonances (blue for nucleus 1, pink for nucleus 2). Interestingly, GRAPE also uses a stimulus at frequency $A_1 + A_2 + \gamma_{\rm n} B_0$ which does not correspond to any natural resonance of the system (purple marker in Fig.~\ref{fig:fft})

We then tried to find a pulse for even shorter evolution times. The shortest evolution time that we could find a pulse for is 2 microseconds with a 0.98 fidelity with a modulated control Hamiltonian in the \textit{x} direction. Note that, when we decreased the total evolution time from 4 microsecond to 2 microseconds, we also decreased the time resolution from 400,000 points to 200,000 points so that the time resolution stayed the same. Overall, GRAPE was consistently able to find one single pulse that can implement the gate without decomposing it into one- and two-qubit gates, and which does not require exceedingly long gate times while taking into account the experimental limitations on the maximum control amplitude.

To understand whether there are any advantages or disadvantages in terms of experimental feasibility of using the $XX$ gate over the decomposed parity check circuit, we took a further step and attempted to implement the decomposed parity check circuit, as defined in Figure~\ref{fig:mpp2circuit}, by individually implementing each one- and two-qubit gate using GRAPE. Surprisingly, while we were able to implement the Hadamard gates within a proper gate evolution time of 0.5 microseconds, the implementation of CNOT gates with high fidelity required a much longer time, around 100 microseconds. As a result, the total timing for the decomposed parity check circuit fell in the $\sim 100$ microsecond range. 
This analysis shows that the system under study -- two nuclear spins hyperfine-coupled to an electron -- is naturally suited to implement a SSPC gate, thanks to the fact that each qubit has a separate resonance frequency, and that the ancilla (the electron) has resonances that each constitute a natural rotation conditional on the state of the two nuclei.
 
\NewDocumentCommand{\Clock}{O{1cm}O{\large}O{cyan}O{}}{%
\def\radius{#1}%
\begin{tikzpicture}[line cap=rect,line width=0.055*\radius]
\filldraw [fill=#3] (0,0) circle [radius=\radius] node {\textbf{#4}};
\foreach \angle [count=\xi] in {60,30,...,-270}
{
  \draw[line width=1pt] (\angle:0.9*\radius) -- (\angle:\radius);
  \node[font=#2] at (\angle:0.68*\radius) {};
}
\foreach \angle in {0,90,180,270}
  \draw[line width=0.04*\radius] (\angle:0.82*\radius) -- (\angle:\radius);
\end{tikzpicture}%
}
\begin{figure}[!htb]
  
  \centering
  \resizebox{\textwidth}{!}{%
  \sidesubfloat[]{
  U = 
  \parbox{2.2cm}{%
    \centering%
    \fcolorbox{red}{white}{\parbox{2.2cm}{%
      \centering
       $e^{-i \Circled[outer color=green]{\gamma_e B_0}  S_z t}$
      \Clock[0.9cm][\footnotesize][green][$\gamma_e B_0$] 
    }}

    $Gate_1 $
  }
  \parbox{2.2cm}{%
    \centering%
    \fcolorbox{red}{white}{\parbox{2.2cm}{%
      \centering
      $e^{-i \Circled[outer color=pink]{\gamma_n B_0}  I_{z1} t}$

      \Clock[0.9cm][\footnotesize][pink][$\gamma_n B_0$]
    }}

    $Gate_2$
  }
  \parbox{2.2cm}{%
    \centering%
    \fcolorbox{red}{white}{\parbox{2.2cm}{%
      \centering
      $e^{-i \Circled[outer color=pink]{\gamma_n B_0}  I_{z2} t}$

      \Clock[0.9cm][\footnotesize][pink][$\gamma_n B_0$]
    }}

    \text{$Gate_3$}
  }
  \parbox{2.2cm}{%
    \centering%
    \fcolorbox{red}{white}{\parbox{2.2cm}{%
      \centering
      $e^{-i \Circled[outer color=orange]{A_1}  S_zI_{z1} t}$
      \Clock[0.9cm][\footnotesize][orange][$A_1$]
    }}

    $Gate_4$
  }
  \parbox{2.2cm}{%
    \centering%
    \fcolorbox{red}{white}{\parbox{2.2cm}{%
      \centering
       $e^{-i \Circled[outer color=yellow]{A_2}  S_zI_{z2} t}$
      \Clock[0.9cm][\footnotesize][yellow][$A_2$]
    }}

    $Gate_5$
  }
  } 
  }\\
  \vspace{0.5cm}
  \setlength{\labelsep}{1 cm}
  \sidesubfloat[]{
  
  \centering
  \leavevmode
  \Large
  \begin{pgfpicture}{0em}{0em}{0em}{0em}
\color{green}
\pgfrect[fill]{\pgfpoint{1.25em}{-.8em}}{\pgfpoint{3.3em}{1.6em}}
\color{pink}
\pgfrect[fill]{\pgfpoint{1.25em}{-2.95em}}{\pgfpoint{3.3em}{1.6em}}
\color{pink}
\pgfrect[fill]{\pgfpoint{1.25em}{-5.1em}}{\pgfpoint{3.3em}{1.6em}}
\color{orange}
\pgfrect[fill]{\pgfpoint{5.5em}{-2.95em}}{\pgfpoint{3.5em}{1.6em}}
\color{yellow}
\pgfrect[fill]{\pgfpoint{9.7em}{-5.2em}}{\pgfpoint{3.6em}{1.6em}}
\end{pgfpicture}
  \Qcircuit @C=1em @R=0.7em {
\lstick{\ket{e}}& \gate{Gate_1} & \ctrl{1} & \ctrl{2} & \qw \\
\lstick{\ket{n_1}}& \gate{Gate_2}\qw & \gate{Gate_4} & \qw & \qw \\
\lstick{\ket{n_2}}& \gate{Gate_3}\qw  & \qw & \gate{Gate_5} & \qw
}
  }
\caption{ a) Clock dynamics of the simplified Ising drift Hamiltonian 
in the laboratory frame. Every term in the time evolution operator of the drift Hamiltonian can be associated with a gate with a specific `clock'. The colors of the clocks differentiate the times for completion of each gates. $Gate_1$ is a single qubit gate on the electron, with clock speed is $\gamma_eB_0$. $Gate_2$ and $Gate_3$ are the single qubit gates on the nuclei, with clock speed $\gamma_nB_0$. The hyperfine interactions in the drift Hamiltonian, $A_1$ and $A_2$, are used to generate the entangling operations $Gate_4$ (between electron and nucleus 1) and $Gate_5$ (between electron and nucleus 2). b) Circuit diagram of the gates depicted in panel a). Note that these are just the gate arising from the drift Hamiltonian. 
} 
\label{fig:clock}
\end{figure}

In the next part of this work, we analytically study a modified drift Hamiltonian to gain an understanding of the appropriate timescale for the SSPC gate and the decomposed parity check circuit. In pursuit of this goal, we performed a thorough investigation of the clock dynamics associated with the drift Hamiltonian in the laboratory frame. To afford an analytical solution, we approximate the drift Hamiltonian as an Ising type modification to Equation~\ref{eq:Hamiltonian}. While the result of GRAPE  
explicitly examines the active pulsing required to realize the unitary for the SSPC gate without loss of generality, here we analytically examine the passive drift Hamiltonian for the Ising Hamiltonian in Equation~\ref{Ising}. 
\begin{equation} \label{Ising}
    H = -\gamma_eB_0S_z - \gamma_nB_0I_{z1} - \gamma_nB_0I_{z2}+ A_1S_zI_{z1}+A_2S_zI_{z2}
\end{equation}
The time evolution operator, $U = e^{iHt}$, can be expanded as
\begin{equation} \label{final_unitary_eq}
\begin{aligned}
  U = \underbrace{e^{-i\mathcircled{\gamma_eB_0}S_Zt}}_\text{$Gate_1$}\underbrace{e^{-i\mathcircled{\gamma_nB_0}I_{z1}t}}_\text{$Gate_2$}\underbrace{e^{-i\mathcircled{\gamma_nB_0}I_{z2}t}}_\textsc{$Gate_3$} 
  \underbrace{e^{i\mathcircled{A_1}S_zI_{z1}t}}_\text{$Gate_4$}\underbrace{e^{i\mathcircled{A_2}S_zI_{z2}t}}_\text{$Gate_5$}
\end{aligned}
\end{equation}
Each of the exponential terms in Equation~\ref{final_unitary_eq}, which are the free evolution times of the drift Hamiltonian, will generate a gate and every gate will have its own clock. These clocks correspond to the time taken for the desired gates. $Gate_1$, $Gate_2$, $Gate_3$, $Gate_4$ and $Gate_5$ are the natural gates which can be obtained from the evolution of the drift Hamiltonian. However, if we want to create specific gates, such as a CZ gate between an electron and the first nucleus using the hyperfine interaction denoted as $A_1$, in other words, if we want $Gate_4$ to function as a CZ gate, then we need to have the control Hamiltonian and pulse the system. In this part of the study, our focus is on understanding the approximate timescale for the SSPC gate and the decomposed parity check circuit that were derived by using GRAPE. When considering the control Hamiltonian, the time required for the SSPC gate and the decomposed parity check circuit will change. However, the relative timescales between the SSPC gate and the decomposed parity check circuit will be comparable for analytical analysis and the numerical analysis. In this part of the study, since we are not using the control Hamiltonian, we do not refer to the gates as CZ gate or $ZZ$ gate. Instead, we refer to them as CZ-like gate and $ZZ$-like gate.
If we want to implement a decomposed \textit{ZZ} parity check circuit as shown in Figure~\ref{fig:mpp2zzcircuit}, we need to implement CZ-like dynamics, since the $ZZ$ parity check circuit can be implemented either by using the CZ gate with Hadamard gates or directly with two CNOT gates. However, if we want to implement an entangling gate between the electron and the first nucleus, we need to use the hyperfine coupling $A_1$. However, the hyperfine coupling $A_2$ is also present and always on. Therefore, we need to find a time that allows us to perform an entangling gate ($Gate_4$) between the electron and the first nucleus and the identity ($Gate_5$) between the electron and the second nucleus, so that the second hyperfine interaction does not disturb the system. To find such a time, we need to consider the clock dynamics of the drift Hamiltonian. The first entangling gate will be between the electron and the first nucleus. A $\pi$ rotation around the Bloch sphere will be a CZ rotation since $e^{i\pi} = -1$. The time for $Gate_4$ to function as a CZ-like gate  will be $\pi/95 \text{ MHz} \approx 33.06
\text{ nanoseconds}$.  
For $Gate_5$ to be the identity operation, it should induce a $2\pi$ rotation, which takes $2\pi/9 \text{ MHz}  \approx 698.13
\text{ nanoseconds}$. However, there will be some error associated with timing, denoted as $t_{\text{error}}$, as achieving the exact duration is not always possible. To find the appropriate time for both gates, we need to solve:
\begin{equation}\label{eq:GAte4}
\begin{aligned}
        ((a \times t_{Gate4}) - (b \times t_{Gate5}))^2 <= t_{\text{error}}^2, a > 0, b > 0, \\ \text{where } a \text{ is an odd integer} \text{ and } b \text{ is an integer}
\end{aligned}
\end{equation}
Here, the timing error, $t_{\text{error}}$ is chosen as 0.01, and consequently, probability of experiencing a Pauli error is equal to the $t_{\text{error}}^2$ is 0.0001. 
In this situation, the evolution of $Gate_4$ should at least make $a =6981$ $\pi$ rotations while the evolution of $Gate_5$ should at least make $b=331$ $2\pi$ rotations so that we can have a perfect CZ-like gate. The total time for a perfect CZ-like gate between the electron and the first nuclei is $33.1*6981 = 231.07~\mu s$

Now, if we want to make a perfect CZ-like gate between the electron and the second nucleus, one $\pi$ rotation for $Gate_5$ to function as a CZ-like is equal to $\pi/9~\text{MHz} \approx 349.06
\text{ nanoseconds}$ and the time for one $2\pi$ rotation for $Gate_4$ to function as identity is equal to $\approx 66.13
$ nanoseconds. If we solve:
\begin{equation}\label{eq:Gate5}
    \begin{aligned}
      ((a\times t_{Gate5}) - (b \times t_{Gate4}))^2 <= t_{\text{error}}^2, a > 0, b > 0, \\ \text{where a is an odd integer and b is an integer}
    \end{aligned}
\end{equation}
the fastest possible time will occur when we take a single $\pi$ rotation time for $Gate_5$ as 349.06 ns and single $2\pi$ rotation time for $Gate_4$ as 66.13 ns. In this case, $a = 887$ is the number of $\pi$ rotations for the evolution of $Gate_5$. Correspondingly, the total time for a clear CZ-like gate for the electron and the second nucleus is equal to $349.06 \times 887 = 309.06~\mu $s. 
As a result, if we want to implement the circuit which is defined in Figure~\ref{fig:mpp2zzcircuit}, the simplified Hamiltonian can naturally achieve that in a range of hundreds $\mu $s. 
In the second part of this analytical calculation, we will analyze the drift Hamiltonian for the  $ZZ$-like gate. For implementing such a gate, both the $Gate_4$ and the $Gate_5$ should function as CZ gates so they should make a $\pi$ rotation. For $Gate_4$ to function as CZ-like, the time for one single $\pi$ rotation is $\pi/95~\text{MHz} = 33.06
\text{ nanoseconds}$ and for $Gate_5$, the time for one single $\pi$ rotation is $\pi/9~\text{MHz} = 349.06
~\text{ nanoseconds}$.  To see if our drift Hamiltonian can have a suitable time for such a gate, we need to solve Equation~\ref{eq:ZZ}.
\begin{equation} \label{eq:ZZ}
    \begin{aligned}
       ((a \times t_{Gate4}) - (b \times t_{Gate5}))^2 <= t_{\text{error}}^2, a > 0, b > 0, \\ \text{where both a and b are an odd integer}
    \end{aligned}
\end{equation}
This yields $a = 1805$ and $b = 171$ and total time for the desired unitary of  
$\approx 60~\mu$s.
 As a result, implementing two separated CZ-like dynamics requires $\approx 540.13~\mu$s in total whereas implementing a $ZZ$-like dynamics requires $\approx 60~\mu$s which is 9 times faster than implementing a traditional parity check circuit. This explanation clarifies why the GRAPE algorithm discovered the SSPC gate approximately 10 times faster than the decomposed parity check circuit. This is because the SSPC gate does not suffer from imbalanced hyperfine interaction, whereas the isolated entangling gate does. The codes concerning this section, encompassing both the $XX$ gate and the decomposed parity check gate, can be found in~\cite{Ustun_Implementing-XX-Single-Step-Parity-Check-Gate-on-Silicon-Spin-System_2023}\cite{Ustun_CNOT_Gate_Implementation_2023}.
\section{Conclusions}\label{sec:results}
In this theory work, we developed a workflow which calculates the perfection rate of the whole parity check with given experimental data.  
We numerically showed how errors are accumulating and causing a decrease of the perfection rate of the parity check circuit with current sources (with single and two qubit gate and $M_{pp1}$) from the experimental GST matrices.
Our findings indicate that relying solely on the fidelity of 1- and 2-qubit gate operations is inadequate in assessing the effectiveness of constructing the fundamental elements of quantum error correction, specifically the parity check circuits.
Then, we introduced new SSPC unitaries which allow us to make parity check measurements in one step. We modified the universal gate set such that the new unitaries are the most natural gate set for QEC. We compared the traditional parity check gate (1- and 2-qubits gates) and the SSPC gate for an \textit{XX} parity check scenario in the presence of noise. 

We provided an experimentally realizable example based on spin qubits in silicon, and used the GRAPE algorithm to find a single pulse capable of implementing the single-step parity check gate within experimental limitations. Furthermore, we implemented the decomposed parity check circuit using GRAPE. We found that the SSPC gate is naturally suited for this system, and outperforms a decomposed gate. 

Many have asked what the ``killer-app" is for a large-scale, error-corrected quantum computer, this is error-correction.  The vast majority of the computation occurring in a large-scale machine is dedicated to correcting its own errors.  As such, fundamental gate sets used at the physical level should be targeted at this application.  We have introduced a complete workflow that allows experimentalists to characterise more complex gate sets and connect outputs of characterisation experiments to commonly used models used in QEC analysis.  This allows quantum hardware engineers to potentially redesign and test fundamental gate libraries, increase performance, and accelerate the realisation of fault-tolerant, error-corrected quantum computers

\section{Data availability statement}
The algorithm for finding the perfection rate and the errors in the context of quantum error correction from experimental data can be found in \cite{Ustun_Error_Analysis_in_2023}.\\
All the code regarding the GRAPE algorithm can be found in \cite{Ustun_Implementing-XX-Single-Step-Parity-Check-Gate-on-Silicon-Spin-System_2023} and in \cite{Ustun_CNOT_Gate_Implementation_2023}
\newpage
\section*{Appendices}
\setcounter{section}{0}
\setcounter{equation}{0}
\setcounter{figure}{0}
\renewcommand{\thesection}{A-\Roman{section}}
\renewcommand{\theequation}{A.\arabic{equation}}
\renewcommand{\thefigure}{A.\arabic{figure}}
\subsection{ Workflow - Understanding the Current Way of Constructing Parity Check Circuit and Its Drawback in the Context of QEC} \label{Mpp1}
The current standard method for constructing multi-body measurements involves applying controlled-gates, like the CNOT (CX) gate for $X$ parity measurement, between data qubits and ancilla qubits, followed by a one-body measurement on the ancilla qubit. The ancilla is crucial in this process, as without it, the measurement result will not provide us with the parity information.
While this works in theory, it is expensive in the lab since each layer of gates added results in a decrease in the total perfection of the quantum circuit. Moreover, we may be able to do better as natural two qubits measurements are possible in some systems.
Here, we provide details on how to derive the total perfection rate of the circuit using stochastic Pauli channels. This calculation assumes that we apply each of the operators just before making the measurement and model every imperfect gate as perfect gates plus some error. 
\subsubsection{Preliminaries}
\begin{enumerate}
    \item Superoperators represent a linear map between initial and final density matrices of the system. They are the mathematical representations which bring one density matrix to another. The unitary evolution of a density matrix is defined by Equation~\ref{supop} where $U$ is the unitary matrix and its dimension is $2^n \times 2^n$, $n$ is the number of qubits, $\rho_i$ is the initial state of the system and $\rho_f$ is the final state of the system after this unitary is applied. 
    \begin{equation} \label{supop}
        \rho_f = \mathcal{E}(\rho_i)= U\rho_i U^{\dagger}
    \end{equation} where $\mathcal{E}$ describes a quantum channel.
    
    Now, there exists a linear representation of $\mathcal{E}$ due to Choi-Jamakowski which brings $\ket{\rho_i}\rangle\mapsto\ket{\rho_f}\rangle$ and its representation based on the vectorization of the density matrix with respect to the column basis:~\cite{GraphicalCAlculus}. 
    \begin{equation}\label{supop2}
       U\otimes U^\dagger\ket{\rho_i}\rangle=\ket{\rho_f}\rangle
    \end{equation}
    Dimensions of the Superoperators are equal to $2^{2n} \times 2^{2n}$ where $n$ is the number of qubits and $ \ket{\cdot} \rangle $ represents vectorization.
    \item Equation~\ref{supop} is the special case where we have applied the unitary $U$ in the absence of environmental noise. However, in the presence of imperfect gates or environmental noise, Equation~\ref{supop} will not be enough to represent the imperfections. In this case, we will need a more general mapping to represent our system. Kraus operators are the most general representation of our system in the presence of noise and they represent both the systematic gate errors and the errors induced by environmental decoherence~\cite{Devitt_2013}. Then, the density matrix can be written as
    \begin{equation} \label{eq:1}
        \rho_f = \mathcal{E}(\rho_i) = \sum_{k=1}^N A_k \rho_i A_k^\dagger,
    \end{equation}
    with
    \begin{equation} \label{eq:2}
        \sum_{k=1}^{N} A_k^\dagger A_k = 1,
    \end{equation} where $N\leq 2^{2n}$ and $\{A_k\}$ are the set of Kraus operators which replace the unitary in Equation~\ref{supop}. The Kraus operators, $\{A_k\}$, are not necessarily unitary operators, hermitian operators or invertible.  
    If $N = 1$, this means that we have the ideal case and this brings us back to the Equation~\ref{supop} where $A_k$ is $U$. Equation~\ref{eq:2} is called the completeness condition if the map is trace-positive, TP. 
    Note that since the quantum channels considered here map the system to itself, we only consider Kraus operators to be square matrices.
    \item $\chi$ - Process matrix\\ 
    When we write Kraus operators in the Pauli basis, in other words, $A_k$ is extended in terms of Pauli operators, then we will have ${A_k = \sum_{j=1}^{d^2} a_{jk} P_j }$ where $ P_j \in P^{\otimes n}$ and ${P=\{I,X,Y,Z\}}$. If we write that in Equation~\ref{eq:1}, then we will have: 
    \begin{equation} \label{eq:3}
        \rho_f = \mathcal{E}(\rho_i) = \sum_{j=1}^{d^2} \chi_{jk} P_j \rho_i P_k
    \end{equation}
    where $\chi_{jk} = \sum_i a_{ij} a_{ik}^*$. The $\chi$ matrix is a complex-valued matrix  with dimension $2^{2n} \times 2^{2n}$ and is a complete map of $\rho_f$. The $\chi$ matrix is also the output of process tomography\cite{Greenbaum_2015}.
    \item The Pauli Transfer Matrix (PTM), which is shown as $R$ in literature~\cite{GraphicalCAlculus}\cite{Greenbaum_2015}\cite{Chow_2012}\cite{korotkov2013error}, is another useful representation of a quantum channel and an outcome of Gate Set Tomography\cite{Greenbaum_2015}\cite{Nielsen_2021}. It is defined by the vectorization of a quantum channel in the Pauli basis and it can be written as: 
    \begin{equation} \label{eq:4}
        (R_{\varepsilon})_{ij} = \frac{1}{d}   \text{Tr}\{P_i \mathcal{E}(P_j)\}
    \end{equation}
    Here, $P_j$'s represents ordering in the strings of Pauli operators, i.e $\{I, X, Y, Z\}$. The PTM, $R$, is the special case of Superoperators where the vectorization is made in the Pauli basis.
    One notable advantage of using the Pauli Transfer Matrix, $R$, is its exclusive utilization of real elements. Furthermore, $R$ facilitates the straightforward determination of whether a quantum operation is trace-preserving or unital. It is also noteworthy that for any Clifford operation, each row and column of R with unit magnitude contains a single non-zero element. 
    The PTM makes it easy to evaluate the result of multiple gates acting in succession. The process matrix does not have this property. This is why the output matrices of Gate Set Tomography(GST) are PTM\cite{Nielsen_2021}. 
    \item The Density Matrix in the context of QEC:

    We shall commence the construction of the density matrix for a single qubit. For the density matrix of multi-qubit systems, we refer the reader to the Appendix. \ref{2-qubit_density}\ref{3-qubit_density}). In this case, the dimensions of Kraus operators will be $2\times2$. Once we found the Kraus operators, they can be written in terms of Pauli matrices as in Equation~\ref{eq:5}, since every complex $2\times2$ matrix can be written by Pauli Matrices
    \footnote{Also every $2^n \times 2^n$ matrix can be written as a linear combination of the tensor products of $n$ Pauli matrices. The total number of independent tensor products of Pauli matrices is $2^{2n}$}. 
    \begin{equation} \label{eq:5}
        A_k = C_0 I + C_1 X + C_2 Y + C_3 Z
    \end{equation}
    If the Kraus operators are in Equation~\ref{eq:1} are decomposed as in Equation~\ref{eq:5}, then we have: 
    \begin{equation} \label{eq:7}
      \begin{aligned}
        \rho_f = \sum\limits_{a\in M} \sum\limits_{b \in M} a \rho_i b^{\dagger}
      \end{aligned}
    \end{equation} where $M=\{C_0I,C_1X,C_2Y,C_3Z\}$.
    
    Suppose the map described by Equation~\ref{eq:7} acts on a qubit that is part of a larger QEC-encoded system. We will now explore the observable effect of this mapping when we run a quantum error correction cycle. For this purpose, let us consider an error model, such as the one given in Equation~\ref{err_model}:
     \begin{equation}\label{err_model}
         \rho_f = p_i\rho + p_x X\rho X + p_{xy} X\rho Y
     \end{equation}
     Here $\rho$ represents a qubit, while $X$ and $Y$ represent the physical error on the qubits in the code block. An ancilla block, represented by the density matrix $\rho_0^{E}$ is coupled to the system and the quantum error correction unitary which is called $U_{QEC}$ is run:
     \begin{equation}
     \begin{aligned}
        U_{QEC}(\rho_f \otimes \rho_0^{E})U_{QEC}^{\dagger} = p_i\rho \otimes \ket{E}\bra{E} 
                                                       +p_x X \rho X \otimes \ket{E_x}\bra{E_x}\\
                                                       +p_{xy} X \rho Y \otimes \ket{E_x}\bra{E_y}
     \end{aligned}
     \end{equation}
     Here, $\ket{E}$, $\ket{E_x}$ and $\ket{E_y}$ represent the three orthogonal syndrome states of the ancilla qubit and they are used the detect errors on the qubits. Now, when the ancilla qubit is measured, the system will be collapsed to one of the two states in Equation~\ref{collapsing}
     \begin{equation}\label{collapsing}
         \begin{aligned}
          \rho_f \Rightarrow \frac{\bra{E} \rho \ket{E} \ket{E}\bra{E}}{\text{Tr}\left(\rho \ket{E}\bra{E}\right)} \text{ and }
           \rho_f \Rightarrow \frac{\bra{E_x} \rho \ket{E_x} \ket{E_x}\bra{E_x}}{\text{Tr}\left(\rho \ket{E_x}\bra{E_x}\right)} 
         \end{aligned}
     \end{equation}
     The off-diagonal term, in this case the $p_{xy} X \rho Y$ term in the error model, is never observed. After the measurement of the ancilla, we have two possible states which are shown in Equation~\ref{measured_state}
     \begin{equation}\label{measured_state}
         \begin{aligned}
             \rho \otimes \ket{E}\bra{E} \text{with probability } p_i  \\
             X \rho X \otimes \ket{E_x}\bra{E_x} \text{with probability } p_x
         \end{aligned}
     \end{equation}
     where $p_i + p_x =1$.
     The off-diagonal term is eliminated because we are about to make a measurement in the $E$, $E_x$ basis and the measurement result will collapse one of the diagonal terms. Although it was written in~\cite{Devitt_2013}, it is once more worth noting that not only the off-diagonal term is eliminated but the final density matrix is collapsed to clean codeword states with bit-flip errors. 
     Here, we do not use the twirling method which is a way to convert the off-diagonal terms of the matrix into the diagonal terms\cite{Cai_2019}\cite{Easwar2008}. In quantum computation, there are some protocols that use twirling, such as purification protocols~\cite{Anwar_2005}\cite{Hou_2014}\cite{Behera_2019}. However, when we write errors in terms of stochastic Pauli channels, we let unitaries evolve until the measurement is done. 
    This is why, when we calculate the effective errors in the context of QEC, we only care about diagonal elements and we ignore off-diagonal elements. Hence, we are finding the diagonal elements of the $\chi$ matrix and we are writing the Kraus operators in terms of these diagonal elements of the $\chi$ matrix. It is true that we are loosing the information by ignoring the off-diagonal terms. In Equation~\ref{eq:7}, once cancelling the off-diagonal elements, we have,
    \begin{equation} \label{eq:8}
        \rho_f= p_i I\rho_i I + p_x X\rho_i X + p_y Y\rho_i Y + p_z Z\rho_i Z
    \end{equation} where $p_i=(|C_0|)^2$, $p_x=(|C_1|)^2$, $p_y=(|C_2|)^2$, $p_z=(|C_3|)^2$.
    This formula represents effective errors in the context of QEC for 1 qubit.  
   As an illustration, in the case of implementing the identity gate, the coefficient $p_i$ serves as an indicator of the gate's efficacy. A value of $p_i=1$ signifies the perfect implementation of the identity gate.  
   The Equation~\ref{eq:8}, can be written for any state and in QEC, the errors in the quantum channel are analyzed after the application of the desired unitary. Because of that, to utilize the Equation~\ref{eq:8} in a more general way, the $\rho_i$ is replaced with $\rho_g$ where $\rho_g = U \rho_i U^\dagger$ and the $U$ is the desired unitary that we want to implement. Then, the Equation~\ref{eq:8}, becomes Equation~\ref{err_channel}:
   \begin{equation} \label{err_channel}
       \rho_f= p_i I\rho_g I + p_x X\rho_g X + p_y Y\rho_g Y + p_z Z\rho_g Z
   \end{equation} 
    In Equation~\ref{err_channel}, the $p_i$ represents the probability of having no error. We define it as the perfection rate. If $p_i=1$, we are in the ideal case with no error and we turn back to Equation~\ref{supop}. For all the other values of $p_i$, we use the Kraus decomposition.
   The coefficients $p_x$, $p_y$ and $p_z$ are,  now, the probability of having \textit{X} errors, the probability of having \textit{Y} errors and the probability of having \textit{Z} errors, respectively. We start with the experimental data, in our case the experimental data is the PTM matrix, and we end up with Equation~\ref{err_channel}. 
   This result is the important conclusion of the section. 
   If we want to determine error rates subsequent to the implementation of a unitary operation within the framework of Quantum Error Correction (QEC), every operation performed in the laboratory will be denoted as Equation~\ref{err_channel}.
   This equation represents a whole process and the coefficient $p_i$ in the equation represents the perfection of the operation that can't be written simply as the fidelity. For example, if we have 99\% fidelity for a specific operation, this does not mean that the coefficient $p_i=0.99$.  
   To find the coefficient $p_i$, we need to do the whole workflow: we first take the experimental process matrices of the system and find the Kraus operators. 
   Subsequently, we then write the Kraus operators in terms of stochastic Pauli channels and determine the value of coefficient $p_i$, which is the perfection rate and it is not equal to fidelity. It is less than the fidelity. Neither the fidelity means the perfection rate, nor the infidelity means the error rate. To finding the error rate, we need information beyond the fidelity~\cite{Sanders_2015}\ref{spp:tutorial}. We give a complete tutorial in the supplementary material regarding this process. If we take Fig.~\ref{fig:1_qubit} as an example and calculate the probability of experiencing specific errors when gates are applied to qubits, the equation that represents a perfection rate of the circuit is Eq.~\ref{eq:12}, for $U_2 = I$. The detailed derivation of Equation~\ref{eq:12} is shown in \ref{P_I}. 
   
    \begin{figure}[H]
      \centering
      \leavevmode
       \LARGE 
         \Qcircuit @C=1.7em @R=2em {
           \ar @[red]  +<0.8 em,0 em>;[]+<0.8 em,-1em>_{\rho_i}   &\gate{U_1} \ar @[red]+<1.8 em,0 em>;[]+<1.8 em,-1em>_{\rho_2}  & \gate{U_2}  \ar @[red]  +<1.8 em,0 em>;[]+<1.8 em,-1em>_{\rho_3} & \qw \\} 
        
        \caption{A one qubit circuit example which will be used to calculate effective errors in the context of QEC. At the beginning, the circuit will have $\rho_i$. After applying a $U_1$ gate, the density matrix will become $\rho_2$ and the final density matrix will be $\rho_3$ which can be written as $\rho_f$. }
        \label{fig:1_qubit}
    \end{figure}
    \begin{equation} \label{eq:12}
     P_I=p_{i2}p_{i1}+p_{x2}p_{x1}+p_{y2}p_{y1}+p_{z2}p_{z1}
    \end{equation}
    The value $P_I$ is the perfection rate for the whole circuit and it represents the final succession rate after all of the accumulated errors in the widget. The important point is that $P_I$ is not simply the product of the individual perfection rates $p_i$. Cancellation terms also contribute to $P_I$, since $X \otimes X = I$, $Y \otimes Y = I$, and $Z \otimes Z = I$. However, this contribution is very small and becomes even smaller for longer circuits with very low error rates. We will call $P_I$ as the perfection rate of the widget. 
    \end{enumerate}
\subsection{Calculating the Final Density Matrix in the Context of QEC}
\begin{figure}[H]
      \centering
      \leavevmode
       \LARGE 
         \Qcircuit @C=1.7em @R=2em {
           \ar @[red]  +<0.8 em,0 em>;[]+<0.8 em,-1em>_{\rho_i}   &\gate{U_1} \ar @[red]+<1.8 em,0 em>;[]+<1.8 em,-1em>_{\rho_2}  & \gate{U_2}  \ar @[red]  +<1.8 em,0 em>;[]+<1.8 em,-1em>_{\rho_3} & \qw \\} 
        
        \caption{One qubit circuit example which will be used to calculate effective errors in the context of QEC. At the beginning, the circuit will have the $\rho_i$, after applying $U_1$ gate, the density matrix will become $\rho_2$ and the final density matrix will be $\rho_3$ after applying $U_3$. }
        \label{fig:1_qubit_}
    \end{figure}
Let'us start with writing $\rho_2$
\begin{equation} \label{rho2}
    \begin{aligned}
      \rho_2 = p_{i1}I U_1 \rho_i U_1^\dagger I + p_{x1}X U_1 \rho_i U_1^\dagger X  + p_{y1}Y U_1 \rho_i U_1^\dagger Y + p_{z1}Z U_1 \rho_i U_1^\dagger Z
    \end{aligned}
\end{equation} where $p_{i1}$ is the perfection rate of $U_1$ gate. $p_x$, $p_y$, and $p_z$ respectively represent the probabilities of having $X$, $Y$, and $Z$ errors in the channel after applying the unitary $U_1$.  The system now in $\rho_2$. Then we apply $U_2$ gates and the system becomes $\rho_3$.
\begin{equation} \label{rho3}
    \begin{aligned}
       \rho_3 = p_{i2}I U_2 \rho_2 U_2^\dagger I + p_{x2}X U_2\rho_2 U_2^\dagger X + p_{y2}Y U_2\rho_2 U_2^\dagger Y + p_{z2}Z U_2 \rho_2 U_2^\dagger Z
    \end{aligned}
\end{equation}
where $p_{i2}$ is the perfection of $U_2$ gate.  $p_{x2}$,$p_{y2}$ and $p_{z2}$ respectively represent the probabilities of having $X$, $Y$, and $Z$ errors in the channel after applying the unitary $U_2$. For the sake of simplicity, we assume that $U_2$ is identity gate. This makes working through the equations much easier since $U_2$ and $X$ commutes. If we put Equation \ref{rho2} in Equation \ref{rho3}, then we will have:
\begin{equation} \label{recursive_Density}
    \begin{aligned}
      \rho_3 =  \dboxed{p_{i2}}I U_2 [\dboxed{p_{i1}}I U_1 \rho_i U_1^\dagger I + p_{x1}X U_1 \rho_i U_1^\dagger X + p_{y1}Y U_1 \rho_i U_1^\dagger Y + p_{z1}Z U_1 \rho_i U_1^\dagger Z] U_2^\dagger I +\\
               \dboxed{p_{x2}}X U_2 [p_{i1}I U_1\rho_i U_1^\dagger I + \dboxed{p_{x1}}X U_1 \rho_i U_1^\dagger X + p_{y1}Y U_1 \rho_i U_1^\dagger Y + p_{z1}Z U_1 \rho_i U_1^\dagger Z] U_2^\dagger X +\\
               \dboxed{p_{y2}}Y U_2 [p_{i1}I U_1 \rho_i I + p_{x1}X U_1 \rho_i U_1^\dagger X + \dboxed{p_{y1}}Y U_1 \rho_i U_1^\dagger Y + p_{z1}Z U_1 \rho_i U_1^\dagger Z]U_2^\dagger Y +\\
               \dboxed{p_{z2}}Z U_2[p_{i1}I U_1 \rho_i U_1^\dagger I + p_{x1}X U_1 \rho_i U_1^\dagger X + p_{y1}Y U_1 \rho_i U_1^\dagger Y + \dboxed{p_{z1}}Z U_1 \rho_i U_1^\dagger Z] U_2^\dagger Z
    \end{aligned}
\end{equation}
Now, the coefficients are inside the dashed lines gives us to total perfection of the circuit after applying operations. As a result we get Equation \ref{P_I}:
\begin{equation}\label{P_I}
\begin{aligned}
  P_I = p_{i2}p_{i1}+p_{x2}p_{x1}+p_{y2}p_{y1}+p_{z2}p_{z1}
\end{aligned}
\end{equation} Here, \( P_I \) denotes the overall perfection rate of the widget—that is, the entire circuit—after the application of all unitaries.  There is one more scenario in which Equation~\ref{P_I} holds: when $U_2$ is not the identity but a depolarizing noise channel is present. For example, if the $U_2$ gate is a Hadamard gate, Equation~\ref{P_I} could include the term $p_{z2}p_{x1}$.
For the accumulated errors, for instance the accumulated x errors, we will have the equation \ref{xerror}:
\begin{equation} \label{xerror}
    \begin{aligned}
      P_X = p_{i2}p_{x1}+p_{x2}p_{i1}+p_{y2}p_{z1}+p_{z2}p_{y1}
    \end{aligned}
\end{equation}
If we had only one step, we would have $\rho_2$ exclusively. In that case, $P_I$ would be equal to $p_{i1}$, and the X error would be represented by $p_x = p_{x1}$. However, as the circuit depth increases, the occurrences of X, Y, and Z errors also increase, leading to a decrease in the fidelity $P_I$. 
In Equation \ref{P_I}, it is important to note that $P_I < p_{i1}$ since errors accumulate within the circuit. Below, we describe our algorithms for extracting the error rates of individual gates from the experimental process matrices and for computing the overall $P_I$ of a quantum circuit.
\subsection{The Algorithm}\label{algorithm}
The algorithm takes experimental process matrices as input and returns the perfection rate of each individual gate, as well as the corresponding Pauli error rates. In this sense, we perform reverse engineering: starting from the quantum channel matrices obtained from experiments on the hardware, we determine the Kraus operators and finally identify the underlying errors by constructing stochastic Pauli channels for each individual gate, rather than assigning gate error rates a priori and simulating them for large-scale applications.

The first step of the algorithm involves identifying the type of process matrices provided. In our group, we use Gate Set Tomography (GST)~\cite{Greenbaum_2015} to characterize the performance of single- and two-qubit gates. For GST, the appropriate matrix representation is the Pauli Transfer Matrix (PTM). 
If the matrix type is not explicitly specified in the code and one attempts to analyze the matrix—for example, to verify its physicality (i.e., whether it is Completely Positive and Trace Preserving, or CPTP)—the check may return a false result simply due to missing type information. Furthermore, misidentifying the matrix type can lead to incorrect conclusions. For instance, if a PTM is mistakenly treated as a \( \chi \)-matrix, even a valid CPTP matrix may be interpreted as invalid because the underlying assumptions about its structure are incorrect. We then find Kraus operator from experimental PTM matrices.

Following the extraction of the Kraus operators, each Kraus operator is expanded in the Pauli basis, and the resulting expressions are used to construct the density matrix in terms of effective errors, as shown in Equation~\ref{eq:8},\ref{err_channel}.

The motivation for the algorithm is to learn the effective Pauli errors of individual gates from experimental process matrices, so that large-scale simulations can then be performed under realistic noise assumptions, directly mapping the gates implemented in a given system to the thresholds associated with a broad class of quantum error correction (QEC) codes. 
The overall perfection value is calculated by including only the contributions from the no-error terms, which have the highest contribution.

The complete code for replicating the results presented in Figure~\ref{fig:xx}, as well as the software for analyzing errors within the context of QEC, can be found in \cite{Ustun_Error_Analysis_in_2023}.
\begin{algorithm}[!htb] \label{algo1}
\renewcommand\thealgorithm{}
\caption{Error Analysis in the Context of Quantum Error Correction} 
\begin{algorithmic}[1]
\phase{Preparation - Finding Kraus Operators from GST}
\State \textbf{Input :} \text{Ideal and experimental GST Matrices}
\State \textbf{Output :} \text{Kraus List}
\State $M_{full} = M_{GST}*M_{GSTerror}$
\State $M_{PTM} = PTM(M_{full}) $ 
\If {$M_{PTM}$ is Complete Positive matrix $== True$}
	 \State $list(M_{Kraus}) = Kraus(M_{PTM})$
	 \If {sum(i), for i in $M_{Kraus} == I$} 
	       \State \textbf{return} $list(M_{Kraus})$
	 \EndIf
\EndIf   
	    
\end{algorithmic} 
\end{algorithm}

\begin{algorithm}[H] \label{algo2}
    \renewcommand\thealgorithm{}
    \caption{Error Analysis in the Context of Quantum Error Correction} 
    \begin{algorithmic}[1]
           \setcounter{phase}{1}
           \phase{Finding errors in the context of Quantum Error Correction for each individual gates}
       	    \State \textbf{Input :} \text{Kraus List}
	        \State \textbf{Output :} \text{Errors}
	        \State i$=$Pauli0,x$=$Pauli1,y$=$Pauli2,z$=$Pauli3,result$=$[],N$=$qubit number
	        \Procedure{Coefficients}{$list(M_{Kraus})$,N}
                \State $paulis =$ list(product([i,x,y,z]),repeat$=$N)   
                \State $spaulis=$ list(product(["i","x","y","z"],repeat$=$N)
        		\For {kindex,kraus in enumerate(krauslist)}:
			        \For {index,pauli in enumerate(paulis)}:
			             \State error = $\frac{1}{2^{N}}$ Trace((kraus*   (TensorProduct(pauli[0],...,pauli[N])))
			        \EndFor
			    \EndFor
			\State \textbf{return} result
			\EndProcedure
			\State d $=$ defaultdict(complex)
            \For {value, name, order in result}: 
                \State  d[name] $+=$ abs(value) * abs(value)
            \EndFor    
	\end{algorithmic} 
\end{algorithm}

\begin{algorithm}[H]
    \renewcommand\thealgorithm{}
    \caption{Error Analysis in the Context of Quantum Error Correction} 
    \begin{algorithmic}[1]
      \setcounter{phase}{2}
      \phase{Computing the Perfection Rate for the each individual step}
            \State \textbf{Input :} 
                            H $=$ errors for H gate in dictionary format, \\ 
                \hspace{1.5cm }\text{    I $=$ errors for I gate in dictionary format,} \\       
                \hspace{1.6cm }cnot $=$ errors for CNOT gate in dictionary format\\
               \hspace{1.5cm } meas $=$ errors for measurement in dictionary format
	        \State \textbf{Output :} Perfection Rate  
	        \State \# For the first part of  2 body parity check circuit which corresponds $H \otimes I \otimes I$
	        \State HII $=$ defaultdict(complex) 
	        \For{nameH, valueH in H.items()}:
	           \For{nameI, valueI in I.items()}:
	              \For{nameI, valueI in I.items()}:
	                 \State name = (*nameH *nameI *nameI)
	                 \State HII[name] = valueH *valueI *valueI
	               \EndFor
	           \EndFor
	        \EndFor  
	        \State \# For the second part of  2 body parity check circuit which corresponds $CNOT \otimes I \otimes I$
	        \State CNOTI $=$ defaultdict(complex)
	        \For{nameCNOT, valuecnot in cnot.items()}:
	           \For{nameI, valueI in I.items()}:
	               \State name = (*nameCNOT *nameI)
	               \State CNOTI[name] = valuecnot *valueI 
	           \EndFor
	        \EndFor
	        \State \# For the third part of  2 body parity check circuit which corresponds non-adjacent CNOT
	        \State ICNOT $=$ defaultdict(complex)
	        \For{nameI, valueI in I.items()}:
	          \For{nameCNOT, valuecnot in cnot.items()}:
	               \State name = ( *nameI *nameCNOT)
	               \State ICNOT[name] =  valueI * valuecnot 
	           \EndFor
	        \EndFor
            \algstore{part1}
            \end{algorithmic}
            \end{algorithm}
           \begin{algorithm}
        \begin{algorithmic}[1]
         \algrestore{part1}
	        \State \# For the last part of  2 body parity check circuit which corresponds measurement gate 
	        \State measII $=$ defaultdict(complex) 
	        \For{nameMeas, valueMeas in meas.items()}:
	           \For{nameI, valueI in I.items()}:
	              \For{nameI, valueI in I.items()}:
	                 \State name = (*nameMeas *nameI *nameI)
	                 \State measII[name] = valueMeas *valueI *valueI
	               \EndFor
	           \EndFor
	        \EndFor  
    \end{algorithmic}
    \begin{algorithmic}
    \State \# \textbf{Overall perfection rate, including only the contributions from having no error terms, which have the highest contribution – (first order approximation, other terms not included.})
    \State $p_I = \text{list}(hii.\text{values()}[0])*\text{list}(cnoti.\text{values()}[0])*\text{list}(icnot.\text{values()}[0])*\text{list}(hii.\text{values()}[0])* \text{list}(meas\_gate.\text{values()}[0])$
\end{algorithmic}
\end{algorithm}

\subsection{Tutorial}\label{spp:tutorial}
Here, we will demonstrate the complete process for a single qubit. We take the circuit in Figure \ref{fig:1_qubitexample} as an example:
    \begin{figure}[H]
      \centering
      \leavevmode
       \LARGE 
         \Qcircuit @C=1.7em @R=2em {
           \ar @[red]  +<0.8 em,0 em>;[]+<0.8 em,-1em>_{\rho_i}   &\gate{\sqrt{X}} \ar @[red]+<2.3 em,0 em>;[]+<2.3 em,-1em>_{\rho_2}  & \gate{I}  \ar @[red]  +<2.3 em,0 em>;[]+<2.3 em,-1em>_{\rho_3} & \qw \\} 
        
        \caption{One qubit circuit example which will be used to calculate effective errors in the context of QEC. At the beginning, the circuit will have the $\rho_i$, after applying $\sqrt{X}$ gate, the density matrix will become $\rho_2$ and the final density matrix will be $\rho_3$ after applying $\sqrt{Y}$. }
        \label{fig:1_qubitexample}
    \end{figure}
We take the data from \cite{Rosty}. The average fidelity of the single qubit operation is 98.4\%. 
The experimental GST matrix for $\sqrt{X}$ gate is \ref{gst}.
\begin{equation} \label{gst}
    \begin{aligned}
     \sqrt{X}_{\text{GST-exp}} = \begin{pmatrix} 
1 & 0 & 0 & 0 \\
8.86 \times 10^{-4} & 0.9864 & 0.01961 & 0.04048 \\ 
0.01433 & 0.01039 & 0.01856 & -0.957\\ 
-0.02782 & -0.03123 & 0.9487 & 0.008478
\end{pmatrix}
    \end{aligned}
\end{equation}
The ideal GST matrix for $\sqrt{X}$ gate is \ref{gst_ideal}.
\begin{equation} \label{gst_ideal}
    \begin{aligned}
      \sqrt{X}_{GST-ideal} = \begin{pmatrix} 1&0&0&0\\0&1&0&0\\0&0&0&-1\\0&0&1&0
      \end{pmatrix}
    \end{aligned}
\end{equation}
Now the complete GST matrix for $\sqrt{X}$ gate will be \ref{fullGST}
\begin{equation} \label{fullGST}
    \begin{aligned}
      \sqrt{X}_{GST} =  \sqrt{X}_{GST-ideal}^\dagger \times \sqrt{X}_{GST-exp}
    \end{aligned}
\end{equation}
\begin{equation} \label{fullgst_matrix}
    \sqrt{X}_{GST} = \begin{pmatrix}
1 & 0 & 0 & 0 \\
8.86 \times 10^{-4} & 0.9864 & 0.01961 & 0.04048 \\
-0.02782 & -0.03123 & 0.9487 & 0.008478 \\
-0.01433 & -0.01039 & -0.01856 & 0.9569
\end{pmatrix}
\end{equation}
We then specify the matrix type in Equation~\ref{fullgst_matrix} as PTM \cite{Greenbaum_2015} \cite{GraphicalCAlculus}\cite{Qiskit} and then, we find the Kraus Operators for $\sqrt{X}$ gate:
\begin{equation}
    \begin{aligned}
     A_{k1}  = \begin{pmatrix}
-0.9828 - i1.522 \times 10^{-17} & 0.01241 - i0.01432 \\
-0.01333 + i5.944 \times 10^{-5} & -0.9898 +i0.02618
\end{pmatrix}
    \end{aligned}
\end{equation}
\begin{equation}
    \begin{aligned}
    A_{k2} = \begin{pmatrix}
0.07268 + i2.130 \times 10^{-16} & 0.09659 - i0.06736 \\
0.1050 - i0.1321 & -0.07130 + i0.004211
\end{pmatrix}
    \end{aligned}
\end{equation}

\begin{equation}
    \begin{aligned}
  A_{k3} = \begin{pmatrix}
9.315 \times 10^{-3} + i4.151 \times 10^{-16} & -2.728 \times 10^{-3} + i1.084 \times 10^{-2} \\
-5.364 \times 10^{-3} - i1.597 \times 10^{-3} & -9.358 \times 10^{-3} + i3.656 \times 10^{-4}
\end{pmatrix}
    \end{aligned}
\end{equation}

\begin{equation}
    \begin{aligned}
     A_{k4} = \begin{pmatrix}
1.140 \times 10^{-3} + i1.923 \times 10^{-12} & -4.041 \times 10^{-4} - i1.773 \times 10^{-3} \\
-3.195 \times 10^{-6} + i1.833 \times 10^{-3} & -1.112 \times 10^{-3} - i2.334 \times 10^{-5}
\end{pmatrix}
    \end{aligned}
\end{equation}
To verify whether we have identified the correct Kraus operators, we need to apply Equation~\ref{eq:2}. After applying Equation~\ref{eq:2}, if we obtain the identity matrix, it indicates that the Kraus operators we have determined are correct.
The subsequent step involves expanding each Kraus operator in terms of Pauli matrices. Therefore, we will have the following expressions:
\begin{equation}\label{Ak1}
    \begin{aligned} 
      A_{k1} = (-0.9863+ i0.01309)I + \\
        (-0.00046 - i0.00713)X + \\
        (0.00719 + i0.01287)Y + \\
        (0.0035 - i0.01309)Z
    \end{aligned}
\end{equation}
\begin{equation}\label{Ak2}
    \begin{aligned}
     A_{k2} = (0.00069+i0.00211)I +\\(0.10081-i0.09971)X + \\(-0.03236-i0.00422)Y+\\(0.07199-i0.00211)Z
    \end{aligned}
\end{equation}
\begin{equation}\label{Ak3}
    \begin{aligned}
     A_{k3} = (-2 \times 10^{-5} + i0.00018)I + \\(-0.00405 + i0.00462)X +\\ (-0.00622 + i0.00132)Y + \\(0.00934 - i0.00018)Z
    \end{aligned}
\end{equation}
\begin{equation}\label{Ak4}
    \begin{aligned}
      A_{k4} =(1 \times 10^{-5} - i1 \times 10^{-5})I + \\(-0.0002 + i3 \times 10^{-5})X +\\ (0.0018 - i0.0002)Y +\\ (0.00113 + i1 \times 10^{-5})Z
    \end{aligned}
\end{equation}
To obtain the total final density matrix for the $\sqrt{X}$ gate, we apply Equation~\ref{eq:7}. We square the absolute value of each coefficient and then sum the coefficients based on their respective groups. As a result, we obtain Equation~\ref{X_final}.
\begin{equation}\label{X_final}
    \begin{aligned}
\rho_{2} =\rho_{\sqrt{X}} = (0.9730)I +\\
                       (0.02019)X \rho_i X +\\
                       (0.001325)Y \rho_i Y+\\
                       (0.005458)Z \rho_i Z
    \end{aligned}
\end{equation}
As seen from Equation~\ref{X_final}, the value of $p_i$ is 0.973, which is not equal to the average fidelity of the 1Q gate operation, which is 98.4\%.
Next, we perform the same process for the $I$ gate, which has an average gate fidelity of 97.5\%.
\begin{equation} \label{gst_y}
    \begin{aligned}
      I_{\text{GST-exp}} = \begin{pmatrix}
1.00000 & 0.00000 & 0.00000 & 0.00000 \\
-0.00018 & 0.93673 & -0.00393 & -0.00473 \\
0.02331 & 0.01384 & 0.93532 & -0.03050 \\
0.01087 & -0.00887 & 0.01742 & 0.98278
\end{pmatrix}
    \end{aligned}
\end{equation}
The ideal GST matrix for $I$ gate is \ref{gst_idealY}.
\begin{equation} \label{gst_idealY}
    \begin{aligned}
      I_{GST-ideal} = \begin{pmatrix} 1& 0&0&0\\0&1&0&0\\0&0&1&0\\0&0&0&1
      \end{pmatrix}
    \end{aligned}
\end{equation}
The final density matrix for $I$ gate is found as Equation \ref{Y_final}.
\begin{equation}\label{Y_final}
    \begin{aligned}
     \rho_3 = \rho_{I} = (0.9637)I +\\
                       (0.0046)X \rho_2 X\\
                       (0.0039)Y \rho_2 Y+\\
                       (0.0276)Z \rho_2 Z
    \end{aligned}
\end{equation}
The perfection rate is found to be $0.9637$, although the fidelity is 97.5\%
As it seen from Equation~\ref{Y_final}, $\rho_3$ has $\rho_2$. To calculate final coefficients of $\rho_3$, we put the Equation~\ref{X_final} in Equation~\ref{Y_final} as it shown in~\ref{recursive_Density}.  $P_I$ coefficient can be then calculating by the Equation~\ref{P_I} and found to be: 
\begin{equation}\label{ex_PI}
\begin{aligned}
     P_I =(0.9637) (0.9730) + \\ (0.0046)(0.02019) + \\(0.0039)(0.001325) + \\(0.0276)(0.00545)\\
\end{aligned}
\end{equation}

\begin{equation}
    P_I = 0.9379
\end{equation}

If the circuit contained only a $\sqrt{X}$ gate, the overall perfection rate would be $0.9730$. Similarly, if it contained only an identity gate, the perfection rate would be $0.9637$. For the two-step circuit, the overall perfection rate is now dominated by the perfection rates of each gate:
\[
(0.9730)(0.9637) = 0.9376,
\]
plus very small contribution from other terms.  

Likewise, if there were only a single $\sqrt{X}$ gate, the $X$ error would be $0.020$. With two gates, the $X$ error is primarily determined by the combination of terms:
\[
(0.02019)(0.9637) + (0.0046)(0.9730) = 0.024,
\]
plus minor contributions from other terms.
\subsection{Density Matrix in Terms of QEC for 2 qubit}\label{2-qubit_density}
\begin{figure}[H]
      \centering
      \leavevmode
       \LARGE 
         \Qcircuit @C=1em @R=0.7em {
           \ar @[red]  +<0.5 em,0 em>;[]+<0.5 em,-4em>_{\rho_i}   & \multigate{1}{U} & \qw &\qw \\
             &\ghost{U} \ar @[red]  +<2.5 em,1.6 em>;[]+<2.5 em, -2.5em>_{\rho_f}&  \qw  &\qw\\
           }     
        \caption{Two Qubits Circuit example with two steps}
        \label{fig:2_qubit}
    \end{figure}
\begin{equation}
 \begin{aligned}
   \rho_f =  p_{ii} I \otimes I U\rho_i U^\dagger  (I\otimes I)^\dagger + p_{xx} X \otimes X U\rho_i U^\dagger (X\otimes X)^\dagger + \\
    p_{yy} Y \otimes YU\rho_i U^\dagger (Y\otimes Y)^\dagger + p_{zz} Z \otimes ZU\rho_i U^\dagger (Z\otimes Z)^\dagger + \\
    p_{ix} I \otimes XU\rho_i U^\dagger (I\otimes X)^\dagger + p_{iy} I \otimes YU\rho_i U^\dagger (I\otimes Y)^\dagger + \\
    p_{iz} I \otimes ZU\rho_i U^\dagger (I\otimes Z)^\dagger + p_{xi} X \otimes IU\rho_i U^\dagger (X\otimes I)^\dagger + \\
    p_{xy} X \otimes YU\rho_i U^\dagger (X\otimes Y)^\dagger + p_{xz} X \otimes ZU\rho_i U^\dagger (X\otimes Z)^\dagger + \\
    p_{yi} Y \otimes IU\rho_i U^\dagger (Y\otimes I)^\dagger + p_{yx} Y \otimes XU\rho_i U^\dagger (Y\otimes X)^\dagger + \\
    p_{yz} Y \otimes ZU\rho_i U^\dagger (Y\otimes Z)^\dagger + p_{zi} Z \otimes IU\rho_i U^\dagger (Z\otimes I)^\dagger + \\
    p_{zx} Z \otimes XU\rho_i U^\dagger (Z\otimes X)^\dagger + p_{zy} Z \otimes YU\rho_i U^\dagger (Z\otimes Y)^\dagger
 \end{aligned}
\end{equation}
\subsection{Density Matrix in Terms of QEC for 3 qubit}\label{3-qubit_density}
 \begin{figure}[H]
      \centering
      \leavevmode
       \LARGE 
         \Qcircuit @C=1em @R=0.7em {
           \ar @[red]  +<0.5 em,0 em>;[]+<0.5 em,-4em>_{\rho}   & \multigate{2}{U} & \qw &\qw \\
             &\ghost{U}& \qw  &\qw\\
             & \ghost{U} & \qw \ar @[red]+<0.3 em,+3.2 em>;[]+<0.3em,-1em>_{\rho_f} \qw & \qw   \\} 
        
        \caption{Three Qubits Circuit example with 2 steps.}
        \label{fig:3_qubit}
    \end{figure}
\begin{equation} \label{DensityMatrix_effectiveerrorsinQEC}
  \begin{aligned}    
  \rho_f = p_{iii} I \otimes I\otimes I U \rho U^\dagger (I\otimes I\otimes I)^\dagger + p_{iix} I \otimes I\otimes XU \rho U^\dagger (I\otimes I\otimes X)^\dagger +\\ p_{iiy} I \otimes I\otimes YU \rho U^\dagger (I\otimes I\otimes Y)^\dagger+p_{iiz} I \otimes I\otimes ZU \rho U^\dagger (I\otimes I\otimes Z)^\dagger+\\
 p_{ixi} I \otimes X \otimes IU \rho U^\dagger (I \otimes X\otimes I)^\dagger + p_{ixx} I \otimes X\otimes XU \rho U^\dagger (I\otimes X\otimes X)^\dagger\\ p_{ixy} I \otimes X \otimes YU \rho U^\dagger (I \otimes X\otimes Y)^\dagger + p_{ixz} I \otimes X \otimes ZU \rho U^\dagger (I \otimes X\otimes Z)^\dagger + \\ p_{iyi} I \otimes Y \otimes IU \rho U^\dagger (I \otimes Y\otimes I)^\dagger + p_{iyx} I \otimes Y \otimes XU \rho U^\dagger (I \otimes Y\otimes X)^\dagger + \\
    p_{iyy} I \otimes Y\otimes YU \rho U^\dagger (I\otimes Y\otimes Y)^\dagger + p_{iyz} I \otimes Y \otimes ZU \rho U^\dagger (Y\otimes Z)^\dagger +\\ p_{izi} I \otimes Z \otimes IU \rho U^\dagger (I \otimes Z\otimes I)^\dagger +
    p_{izx} Z \otimes XU \rho U^\dagger (Z\otimes X)^\dagger +\\ p_{izy} I \otimes Z \otimes YU \rho U^\dagger (I \otimes Z\otimes Y)^\dagger+p_{izz} I \otimes Z\otimes ZU \rho U^\dagger (I\otimes Z\otimes Z)^\dagger +\\ p_{xii} X \otimes I\otimes IU \rho U^\dagger (X\otimes I\otimes I)^\dagger + p_{xix} X \otimes I\otimes XU \rho U^\dagger (X\otimes I\otimes X)^\dagger+\\ p_{xiy} X \otimes I\otimes YU \rho U^\dagger (X\otimes I\otimes Y)^\dagger + p_{xiz} X \otimes I\otimes ZU \rho U^\dagger (X\otimes I\otimes Z)^\dagger +\\ 
    p_{xxi} X \otimes X\otimes IU \rho U^\dagger (X\otimes X\otimes I)^\dagger + p_{xxx} X \otimes X\otimes XU \rho U^\dagger(X\otimes X\otimes X)^\dagger+\\ p_{xxy} X \otimes X\otimes YU \rho U^\dagger (X\otimes X\otimes Y)^\dagger + p_{xxz} X \otimes X\otimes ZU \rho U^\dagger (X\otimes X\otimes Z)^\dagger+\\
     p_{xyi} X \otimes Y \otimes IU \rho U^\dagger (X \otimes Y\otimes I)^\dagger + p_{xyx} X \otimes Y \otimes XU \rho U^\dagger (X \otimes Y\otimes X)^\dagger + \\ p_{xyy}  X \otimes Y \otimes Y\rho (X \otimes Y\otimes Y)^\dagger + p_{xyz} X \otimes Y \otimes ZU \rho U^\dagger (X \otimes Y\otimes Z)^\dagger + \\
     p_{xzi} X \otimes Z\otimes IU \rho U^\dagger (X\otimes Z\otimes I)^\dagger + p_{xzx} X \otimes Z\otimes XU \rho U^\dagger(X\otimes Z\otimes X)^\dagger+\\ p_{xzy} X \otimes Z\otimes YU \rho U^\dagger (X\otimes Z\otimes Y)^\dagger + p_{xzz} X \otimes Z\otimes ZU \rho U^\dagger (X\otimes Z\otimes Z)^\dagger +\\ 
     p_{yii} Y \otimes I\otimes IU \rho U^\dagger (Y\otimes I\otimes I)^\dagger + p_{yix} Y \otimes I\otimes XU \rho U^\dagger (Y\otimes I\otimes X)^\dagger +\\ p_{yiy} Y \otimes I\otimes YU \rho U^\dagger (Y\otimes I\otimes Y)^\dagger+p_{yiz} Y \otimes I\otimes ZU \rho U^\dagger (Y\otimes I\otimes Z)^\dagger+\\
     p_{yxi} Y \otimes X\otimes IU \rho U^\dagger (Y\otimes X\otimes I)^\dagger + p_{yxx} Y \otimes X\otimes XU \rho U^\dagger (Y\otimes X\otimes X)^\dagger +\\ p_{yxy} Y \otimes X\otimes YU \rho U^\dagger (Y\otimes X\otimes Y)^\dagger+p_{yxz} Y \otimes X\otimes ZU \rho U^\dagger (Y\otimes X\otimes Z)^\dagger+\\
     p_{yyi} Y \otimes Y\otimes IU \rho U^\dagger (Y\otimes Y\otimes I)^\dagger + p_{yyx} Y \otimes Y\otimes XU \rho U^\dagger (Y\otimes Y\otimes X)^\dagger +\\ p_{yyy} Y \otimes Y\otimes YU \rho U^\dagger (Y\otimes Y\otimes Y)^\dagger+p_{yyz} Y \otimes Y\otimes ZU \rho U^\dagger (Y\otimes Y\otimes Z)^\dagger+\\
     p_{yzi} Y \otimes Z\otimes IU \rho U^\dagger (Y\otimes Z\otimes I)^\dagger + p_{yzx} Y \otimes Z\otimes XU \rho U^\dagger(Y\otimes Z\otimes X)^\dagger +\\ p_{yzy} Y \otimes Z\otimes YU \rho U^\dagger (Y\otimes Z\otimes Y)^\dagger+p_{yzz} Y \otimes Z\otimes ZU \rho U^\dagger (Y\otimes Z\otimes Z)^\dagger+\\
     p_{zii} Z \otimes I\otimes IU \rho U^\dagger (Z\otimes I\otimes I)^\dagger + p_{zix} Z \otimes I\otimes XU \rho U^\dagger (Z\otimes I\otimes X)^\dagger +\\ p_{ziy} Z \otimes I\otimes YU \rho U^\dagger (Z\otimes I\otimes Y)^\dagger+p_{ziz} Z \otimes I\otimes ZU \rho U^\dagger (Z\otimes I\otimes Z)^\dagger+\\
     p_{zxi} Z \otimes X\otimes IU \rho U^\dagger (Z\otimes X\otimes I)^\dagger + p_{zxx} Z \otimes X\otimes XU \rho U^\dagger (Z\otimes X\otimes X)^\dagger +\\ p_{zxy} Z \otimes X\otimes YU \rho U^\dagger (Z\otimes X\otimes Y)^\dagger+p_{zxz} Y \otimes X\otimes ZU \rho U^\dagger (Z\otimes X\otimes Z)^\dagger+\\
     p_{zyi} Z \otimes Y\otimes IU \rho U^\dagger (Z\otimes Y\otimes I)^\dagger + p_{zyx} Z \otimes Y\otimes XU \rho U^\dagger (Z\otimes Y\otimes X)^\dagger +\\ p_{zyy} Z \otimes Y\otimes YU \rho U^\dagger (Z\otimes Y\otimes Y)^\dagger+p_{zyz} Z \otimes Y\otimes ZU \rho U^\dagger (Z\otimes Y\otimes Z)^\dagger+\\
     p_{zzi} Z \otimes Z\otimes IU \rho U^\dagger (Z\otimes Z\otimes I)^\dagger + p_{zzx} Z \otimes Z\otimes XU \rho U^\dagger (Z\otimes Z\otimes X)^\dagger +\\ p_{zzy} Z \otimes Z\otimes YU \rho U^\dagger (Z\otimes Z\otimes Y)^\dagger+p_{zzz} Z \otimes Z\otimes ZU \rho U^\dagger (Z\otimes Z\otimes Z)^\dagger+\\
 \end{aligned}
\end{equation}
\subsection{$P_I$ for the identity channel in a three-qubit, five-step circuit from $64^5$ terms}
Equation~\ref{P_I} can be extended to a three-qubit, five-step circuit as follows:
\begin{equation} \label{DominatedTerm}
    \begin{aligned}
      (P_I)=(p_{4iii})(p_{3iii})(p_{2iii})(p_{1iii})+p_{4iix}p_{3iix}p_{2iix}p_{iix}+\\
          p_{4iiy}p_{3iiy}p_{2iiy}p_{iiy}+p_{4iiz}p_{3iiz}p_{2iiz}p_{iiz}+\\
          p_{4ixi}p_{3ixi}p_{2ixi}p_{ixi}+p_{4ixx}p_{3ixx}p_{2ixx}p_{ixx}+\\
          p_{4ixy}p_{3ixy}p_{2ixy}p_{ixy}+p_{4ixz}p_{3ixz}p_{2ixz}p_{ixz}+\\
          p_{4iyi}p_{3iyi}p_{2iyi}p_{iyi}+p_{4iyx}p_{2iyx}p_{2iyx}p_{iyx}+\\
          p_{4iyy}p_{3iyy}p_{2iyy}p_{iyy}+p_{4iyz}p_{2iyz}p_{2iyz}p_{iyz}+\\
          p_{4iyz}p_{3iyz}p_{2iyz}p_{iyz}+p_{4izi}p_{3izi}p_{2izi}p_{izi}+\\
          p_{4izx}p_{3izx}p_{2izx}p_{izx}+p_{4izy}p_{3izy}p_{2izy}p_{izy}+\\
          p_{4izy}p_{3izy}p_{2izy}p_{izy}+p_{4izz}p_{3izz}p_{2izz}p_{izz}+\\
          p_{4xii}p_{3xii}p_{2xii}p_{xii}+p_{4xix}p_{3xix}p_{2xix}p_{xix}+\\
          p_{4xiy}p_{3xiy}p_{2xiy}p_{xiy}+p_{4xiz}p_{3xiz}p_{2xiz}p_{xiz}+\\
          p_{4xxi}p_{3xxi}p_{2xxi}p_{xxi}+p_{4xxx}p_{3xxx}p_{2xxx}p_{xxx}+\\
          p_{4xxy}p_{3xxy}p_{2xxy}p_{xxy}+p_{4xxz}p_{3xxz}p_{2xxz}p_{xxz}+\\
          p_{4xxz}p_{3xxz}p_{2xxz}p_{xxz}+p_{4xyi}p_{3xyi}p_{2xyi}p_{xyi}+\\
          p_{4xyx}p_{3xyx}p_{2xyx}p_{xyx}+p_{4xzi}p_{3xzi}p_{2xzi}p_{xzi}+\\
          p_{4xzx}p_{3xzx}p_{2xzx}p_{xzx}+p_{4xzy}p_{3xzy}p_{2xzy}p_{xzy}+\\
          p_{4xzz}p_{3xzz}p_{2xzz}p_{xzz}+p_{4yii}p_{3yii}p_{2yii}p_{yii}+\\
          p_{4yix}p_{3yix}p_{2yix}p_{yix}+p_{4yiy}p_{3yiy}p_{2yiy}p_{yiy}+\\
          p_{4yiz}p_{3yiz}p_{2yiz}p_{yiz}+p_{4yxi}p_{3yxi}p_{2yxi}p_{yxi}+\\
          p_{4yxx}p_{3yxx}p_{2yxx}p_{yxx}+p_{4yxy}p_{3yxy}p_{2yxy}p_{yxy}+\\
          p_{4yxz}p_{3yxz}p_{2yxz}p_{yxz}+p_{4yyi}p_{3yyi}p_{2yyi}p_{yyi}+\\
          p_{4yyx}p_{3yyx}p_{2yyx}p_{yyx}+p_{4yyy}p_{3yyy}p_{2yyy}p_{yyy}+\\
          p_{4yyz}p_{3yyz}p_{2yyz}p_{yyz}+p_{4yzi}p_{3yzi}p_{2yzi}p_{yzi}+\\
          p_{4yzx}p_{3yzx}p_{2yzx}p_{yzx}+p_{4yzz}p_{3yzz}p_{2yzz}p_{yzz}+\\
          p_{4zii}p_{3zii}p_{2zii}p_{zii}+p_{4zix}p_{3zix}p_{2zix}p_{zix}+\\
          p_{4ziy}p_{3ziy}p_{2ziy}p_{ziy}+p_{4ziz}p_{3ziz}p_{2ziz}p_{ziz}+\\
          p_{4zxi}p_{3zxi}p_{2zxi}p_{zxi}+p_{4zxx}p_{3zxx}p_{2zxx}p_{zxx}+\\
          p_{4zxy}p_{3zxy}p_{2zxy}p_{zxy}+p_{4zxz}p_{3zxz}p_{2zxz}p_{zxz}+\\
          p_{4zyi}p_{3zyi}p_{2zyi}p_{zyi}+p_{4zyx}p_{3zyx}p_{2zyx}p_{zyx}+\\
          p_{4zyy}p_{3zyy}p_{2zyy}p_{zyy}+p_{4zyz}p_{3zyz}p_{2zyz}p_{zyz}+\\
          p_{4zzi}p_{3zzi}p_{2zzi}p_{zzi}+p_{4zzx}p_{3zzx}p_{2zzx}p_{zzx}+\\
          p_{4zzy}p_{3zzy}p_{2zzy}p_{zzy}+p_{4zzz}p_{3zzz}p_{2zzz}p_{zzz}+\\
    \end{aligned}
\end{equation}
where the greatest contribution arises from the first term, with only negligible contributions from the other terms.
\printbibliography 

\end{document}